\documentclass[12pt]{iopart}
\usepackage{iopams}
\usepackage{graphicx}
\usepackage{caption}
\usepackage{subcaption}
\usepackage{hyperref}
\usepackage{hypernat}

\begin{document}

\title[Random walks generated by functions of Laplacian matrices]{Random walks with long-range steps generated by functions of Laplacian matrices}
\author{A. P. Riascos${}^\dagger$}  
\address{Department of Civil Engineering, Universidad Mariana,  San Juan de Pasto, Colombia}
\author{T.M. Michelitsch${}^\star$, B.A. Collet} 
\address{Sorbonne Universit\'e, Institut Jean le Rond d'Alembert, CNRS UMR 7190,\\ 4 place Jussieu, 75252 Paris cedex 05, France}
\author{A. F. Nowakowski and F.C.G.A. Nicolleau}
\address{Sheffield Fluid Mechanics Group, Department of Mechanical Engineering, University of Sheffield, Mappin Street, Sheffield S1 3JD, United Kingdom}
\eads{${}^\dagger$aaappprrr@gmail.com, ${}^\star$michel@lmm.jussieu.fr}
\date{\today}

\begin{abstract}
In this paper, we explore different Markovian random walk strategies on networks with transition probabilities between nodes defined in terms of functions of the Laplacian matrix. We generalize random walk strategies with local information in the Laplacian matrix, that describes the connections of a network, to a dynamics determined by functions of this matrix. The resulting processes are non-local allowing transitions of the random walker 
from one node to nodes beyond its nearest neighbors. We find that only two types of Laplacian functions are admissible with distinct behaviors for long-range steps in the infinite network limit: type (i) functions generate Brownian motions, type (ii) functions L\'evy flights. For this asymptotic long-range step behavior only the lowest non-vanishing order of the Laplacian function is relevant, namely first order for type (i), and fractional order for type (ii) functions.
\\[2mm]
In the first part, we discuss spectral properties of the Laplacian matrix and a series of relations that are maintained by a particular type of functions that allow to define random walks on any type of undirected connected networks. Once described general properties, we explore characteristics of random walk strategies that emerge from particular cases with functions defined in terms of exponentials, logarithms and powers of the Laplacian as well as relations  of these dynamics with non-local strategies like L\'evy flights and fractional transport. Finally, we analyze the global capacity of these random walk strategies to explore networks like lattices and trees and different types of random and complex networks. 
\end{abstract}
%
\newpage
\section{Introduction}
The study of dynamical processes taking place on networks have had a significant impact in different fields of science and engineering, leading to important applications in the context of physics, biology, social and computer systems among many others \cite{VespiBook}. In particular, the dynamics of  a random walker that hops visiting the nodes of the network following different strategies is a problem of utmost importance due to connections with a vast of interdisciplinary topics like the ranking of the Internet \cite{GooglePR1998}, transport on networks \cite{GraphsFlow}, the modeling of human mobility in urban settlements \cite{RiascosMateosPlos2017}, recommending links in social networks \cite{Backstrom2011}, chemical reactions \cite{Chemical}, digital image processing \cite{GradyRW2006}, algorithms for extracting useful information from data \cite{Tremblay2014,FoussBook2016}, just to mention a few examples of the implementations.
\\[2mm]
Different types of random walk strategies on networks have been introduced in the last decades, some of them only require local information of  each node and in this way the walker moves form one node to one of its nearest neighbors \cite{montroll:167,HughesBook,NohRieger}, whereas in other cases, the total architecture of the network is available and non-local strategies use all this information to define long-range transitions on the the network \cite{RiascosMateosLF2012,Estrada2017}. The simplest case (but not least important) for a random walker on a network is the normal random walk for which the walker can jump from one node to any of its nearest neighbors with equal probability. This paradigmatic case has been explored in detail for different structures that range from regular to random and complex networks \cite{montroll:167,Weiss,HughesBook,NohRieger,GMFPTBenichou,MasudaPhysRep2017}.  
\\[2mm]
On the other hand, there are different cases for which the information on the total structure of the network is implemented in the definition of a dynamical process. This is the case of the non-local random search strategy introduced in the PageRank algorithm \cite{GooglePR1998,MasudaPhysRep2017}, the L\'evy flights on networks \cite{RiascosMateosLF2012,Zhao2014212,Huang2014132,Weng2015,Weng2016}, L\'evy random walks on multiplex networks \cite{Guo2016}, the fractional diffusion on networks and lattices \cite{RiascosMateosFD2014,RiascosMateosFD2015,Michelitsch2016,Michelitsch2016Chaos,Michelitsch2017PhysA,deNigris2016,DeNigris2017,Michelitsch2017PhysARecurrence}, the quantum transport on networks \cite{RiascosMateos2015}, the graph-based semi-supervised learning \cite{SdeNigris2017}, the dynamics of agents moving visiting specific sites in  a city \cite{RiascosMateosPlos2017} and different strategies in the context of the random multi-hopper model  \cite{Estrada2017}. The study and possible applications of non-local dynamical processes on networks is a relatively new field that opens questions related with the exploration of the effects that non-locality introduces and the search of global quantities that allow to compare the performance of non-local against local dynamics. 
\\[2mm]
In this paper, we explore different  random walk strategies on networks with transition probabilities defined in terms of a family of functions of the Laplacian matrix that describes the network. In the first part, we discuss general properties of the eigenvalues and eigenvectors of the Laplacian matrix of connected undirected networks and how particular characteristics are preserved by a family of functions that are suitable to define random walks on any type of undirected connected networks. The formalism introduced allows to generalize different results and techniques developed in the context of the fractional Laplacian of a graph \cite{RiascosMateosFD2014}.  Once discussed the general case, we explore characteristics of random walk strategies that emerge from particular functions defined in terms of exponentials, logarithms and powers of the Laplacian as well as relations of these dynamics with non-local strategies like L\'evy flights and the fractional transport. Finally, we analyze the Kemeny 
constant and a global time that measures the capacity of different types of random walk strategies to explore networks like lattices, trees and random networks generated with the Erd\H{o}s–R\'enyi, Watts-Strogatz and Barab\'asi-Albert models.

\section{Eigenvalues and eigenvectors of the Laplacian matrix}
The spectral analysis of matrices associated to networks reveal structural properties and is an important  tool in the the study of dynamical processes taking place on networks \cite{VanMieghem,GodsilBook}. In this section we present some basic definitions and results about the Laplacian matrix of a simple undirected graph that describes the topology of a network and properties related to the eigenvalues and eigenvectors of this matrix.
\subsection{Laplacian matrix}
We consider undirected simple connected networks with $N$ nodes $i=1,\ldots ,N$. The topology of the network is described by an adjacency matrix $\mathbf{A}$ with elements $A_{ij}=A_{ji}=1$ if there is an edge (or link) between the nodes $i$ and $j$ and $A_{ij}=0$ otherwise; in particular, $A_{ii}=0$ avoiding links that connect a node with itself. In terms of the elements of the adjacency matrix, the degree $k_i$ of the node $i$ is the number of neighbors that this node has and is given by $k_i=\sum_{l=1}^N A_{il}$. Now, by using this notation,  the Laplacian matrix $\mathbf{L}$ of a network with $N$ nodes is a symmetric $N\times N $ matrix with elements $L_{ij}$ given by \cite{NewmanBook,GodsilBook}
\begin{equation}\label{DefLapl}
L_{ij}=k_{i}\delta_{ij}-A_{ij}
\end{equation}
for $i,j=1,2,\ldots ,N$ and, where $\delta_{ij}$ denotes the Kronecker delta. In matrical representation we have $\mathbf{L}=\mathbf{K}-\mathbf{A}$, where we denote with $\mathbf{K}$ to the diagonal matrix with the node degrees $k_1,k_2,\ldots, k_N$ in the diagonal entries. In addition, from Eq. (\ref{DefLapl}) we observe that non-diagonal elements of $\mathbf{L}$ are negative or null, then $L_{ij}\leq 0$ for $i\neq j$.
\\[2mm]
One of the most important properties of the Laplacian matrix is that this matrix defines a quadratic form. In this way, for an arbitrary column vector $\mathbf{x}$ in $\mathbb{R}^N$ with components $x_1,x_2,\ldots ,x_N$, we have \cite{GodsilBook}
\begin{equation}\label{positiveL}
\mathbf{x}^T \, \mathbf{L}\, \mathbf{x}=\sum_{(i,j)\in \mathcal{E}}(x_i-x_j)^2\geq 0,
\end{equation}
where $\mathcal{E}$ denotes the set of edges of the network and the row vector $\mathbf{x}^T$ is the transpose of  $\mathbf{x}$. The result in Eq.  (\ref{positiveL}) is obtained from the following relation
\begin{eqnarray*}
&\frac{1}{2}\sum_{i=1}^N\sum_{j=1}^N A_{ij}(x_i-x_j)^2 =\frac{1}{2} \sum_{i=1}^N\sum_{j=1}^N A_{ij}\left(x_i^2+x_j^2-2x_i x_j \right) \\
&= \frac{1}{2}\left( \sum_{i=1}^N 2 x_i^2 \underbrace{\sum_{j=1}^N A_{ij}}_{k_i} -2 \sum_{i=1}^N\sum_{j=1}^N A_{ij}x_ix_j \right),
\end{eqnarray*}
where the prefactor $2$ in the first term of last expression comes into play by using $A_{ij}=A_{ij}$, then this equation writes
\begin{eqnarray*}
&\frac{1}{2}\sum_{i=1}^N\sum_{j=1}^N A_{ij}(x_i-x_j)^2 = \sum_{i=1}^N k_i x_i^2 - \sum_{i=1}^N\sum_{j=1}^N A_{ij}x_i x_j \\
&= \sum_{i=1}^N\sum_{j=1}^N(k_i\delta_{ij}-A_{ij})x_i x_j = \sum_{i=1}^N\sum_{j=1}^N L_{ij}x_i x_j.
\end{eqnarray*}
We can hence write Eq.  (\ref{positiveL}) in the form of the last relation. The result in Eq. (\ref{positiveL}) implies that $\mathbf{L}$ is a positive semidefinite matrix and therefore its eigenvalues are all non-negative \cite{GodsilBook}.
\\[2mm]
The Laplacian matrix contains all the information associated to the topology of a network and in this way is fundamental in the study of its  characteristics as well as the analysis of dynamical processes taking place on networks. Diverse works have addressed this topic; in particular, the classic books of Godsil \cite{GodsilBook}, Chung \cite{ChungBook} and the recent work of Van Mieghem \cite{VanMieghem}, review different aspects of algebraic graph theory and the spectra of networks. In the context of dynamical processes, properties of the Laplacian matrix have been explored in studies about synchronization  and its relation with structural properties of networks \cite{Arenas_Synch}, random walks and diffusion on networks and lattices \cite{Lovasz1996, RandomWalksPenmanBook,LawlerBook}, continuous-time quantum walks \cite{MulkenPR502}, advanced techniques and algorithms for extracting useful information from network data \cite{FoussBook2016}, among many other processes \cite{VespiBook}. 
\subsection{General properties of the Laplacian spectra}
\label{SectLapProp}
In this part we present some general aspects related with the spectrum of the Laplacian matrix and their respective eigenvectors. Since $\mathbf{L}$ is a symmetric matrix, using the Gram-Schmidt orthonormalization of the eigenvectors of $\mathbf{L}$, we obtain a set of eigenvectors $\{\left|\Psi_j\right\rangle\}_{j=1}^N$  that satisfy the eigenvalue equation
\begin{equation}
\mathbf{L}\left|\Psi_j\right\rangle=\mu_j\left|\Psi_j\right\rangle, \qquad j=1,\ldots,N.
\end{equation}
The eigenvalues of the Laplacian matrix are $\{ \mu_j \}_{j=1}^N$ and as a direct consequence of the symmetry and the result in Eq. (\ref{positiveL}), the eigenvalues  of $\mathbf{L}$ are real and non-negative. In the following, at least we specify the contrary, the set of eigenvalues is sorted in increasing order, then
\begin{equation}\label{PositiveMu}
0\leq \mu_1\leq \mu_2 \leq \mu_3 \leq \ldots \leq \mu_N.
\end{equation}
For the set of eigenvectors, we have the orthonormalization condition
\begin{equation}
\left\langle\Psi_i|\Psi_j\right\rangle=\delta_{ij}.
\end{equation}
In addition, this set of eigenvectors satisfies the completeness relation
\begin{equation}
\sum_{l=1}^N \left |\Psi_l\right\rangle\left\langle\Psi_l \right|=\mathbb{I}.
\end{equation}
where $\mathbb{I}$ denotes the $N\times N $ identity matrix. Once introduced the basic notation for the eigenvalues and eigenvectors of $\mathbf{L}$, the spectral form of the Laplacian is
\begin{equation}\label{FunGLap}
\mathbf{L}=\sum_{m=1}^N \mu_m \left |\Psi_m\right\rangle\left\langle\Psi_m \right|.
\end{equation}
And, from this result we obtain for the trace (denoted as $\Tr(\ldots)$) of the Laplacian matrix
\begin{eqnarray*}
\Tr(\mathbf{L})&=\sum_{i=1}^N L_{ii}=\sum_{i=1}^N\sum_{m=1}^N \mu_m \langle i |\Psi_m\rangle\langle\Psi_m|i \rangle\\
&=\sum_{m=1}^N \mu_m \langle\Psi_m |\left[\sum_{i=1}^N |i \rangle \langle i |\right] |\Psi_m\rangle =\sum_{m=1}^N \mu_m \langle\Psi_m |\Psi_m\rangle =\sum_{m=1}^N \mu_m .
\end{eqnarray*}
However, $\Tr(\mathbf{L})=\sum_{i=1}^N k_i$, result calculated directly from Eq. (\ref{DefLapl}). Consequently we have an invariant that relates the topology of the network with the eigenvalues of the Laplacian matrix
\begin{equation}
 \langle k \rangle= \frac{1}{N} \sum_{m=1}^N \mu_m,
\end{equation}
where $\langle k \rangle=\frac{1}{N} \sum_{i=1}^N k_i$ is the average degree of the network.
\\[2mm]
%
%
%
%
%
%
%
On the other hand, from the definition in Eq. (\ref{DefLapl}), the Laplacian satisfies
\begin{equation}\label{SumLprop}
 \sum_{j=1}^N L_{ij}=\sum_{j=1}^N k_i\delta_{ij}-\sum_{j=1}^N A_{ij}=k_i-k_i=0.
\end{equation}
This relation in the elements of the Laplacian matrix introduces a restriction in the smallest eigenvalue of $\mathbf{L}$ and the associated eigenvector. In fact, the result in Eq. (\ref{SumLprop}) requires
\begin{equation}\label{EigVectorL1}
|\Psi_1\rangle=\frac{1}{\sqrt{N}}
\left[
\begin{array}{c}
 1 \\
 1 \\
\ldots \\
 1 \\
\end{array}
\right]
\end{equation}
and the corresponding eigenvalue is $\mu_1=0$. This is the lower bound of the Laplacian spectrum. In addition, the multiplicity of $\mu_1=0$, i.e. the number of eigenvalues with this value, is related to the connectivity of the network. In general, the multiplicity of the smallest eigenvalue  of the Laplacian $\mathbf{L}$ is equal to the number of independent connected components in the network \cite{VanMieghem}. For a connected graph $\mu_1=0$ is unique, thus
\begin{equation}
0<\mu_2 \leq \mu_3 \ldots \leq \mu_N.
\end{equation}
The second smallest eigenvalue $\mu_2$ gives us information about the connectivity of the graph. This quantity has been extensively explored in the context of partitioning of graphs \cite{LaplacianBook2007}.  Fiedler called $\mu_2$ the {\it algebraic connectivity of a graph} \cite{Fiedler1973}, and the corresponding eigenvector $|\Psi_2\rangle$ is known as Fiedler vector \cite{LaplacianBook2007}.  For connected networks, the second smallest eigenvalue $\mu_2$ satisfies \cite{VanMieghem}
\begin{equation}
0<\mu_2\leq \frac{N}{N-1} k_{\mathrm{min}},
\end{equation}
where $k_{\mathrm{min}}$ denotes the minimum degree encountered in the network.
The two smallest eigenvalues $\mu_1$ and $\mu_2$ of $\mathbf{L}$ are important in the study of dynamical properties of processes defined in terms of the Laplacian, in a similar way the largest eigenvalue $\mu_N$ allows us to study particular asymptotic limits. The eigenvalue $\mu_N$ satisfies the inequality \cite{VanMieghem}
\begin{equation}\label{mumax}
k_{\mathrm{max}}+1\leq \mu_N \leq \max\{N,2k_{\mathrm{max}}\},
\end{equation}
in this relation $k_{\mathrm{max}}$ is the largest degree of the nodes in the network. 

\section{Functions that preserve the Laplacian structure}
\label{FunctionSection}
The objective of this section is to explore functions $g(\mathbf{L})$ of the Laplacian matrix. From these functions we can obtain new matrices that combine all the information of a  network and for which emerge non-local correlations. We can take advantage of this non-locality to define quantities that describe the whole graph or to introduce new dynamical processes on networks.
\subsection{Function $g(\mathbf{L})$ and general conditions}
For a well defined function $g(x)$ with $x\in \mathbb{R}$, the matrix $g(\mathbf{L})$ can be obtained by using the series expansion $g(x)=\sum_{n=0}^\infty c_n x^n$ or in terms of the spectral form of the Laplacian $\mathbf{L}$. For the second option, we have
\begin{equation}\label{FunGLap}
g(\mathbf{L})=\sum_{m=1}^N g(\mu_m) |\Psi_m\rangle\langle\Psi_m| \, .
\end{equation}
In the following we denote as $g_{ij}(\mathbf{L})$ to the $i,j$ element of the matrix $g(\mathbf{L})$, this notation is maintained for different functions of matrices explored in the rest of this section.
\\[2mm]
Although the result in Eq. (\ref{FunGLap}) allows to calculate general functions of the Laplacian, we are only interested in particular functions that preserve the {\it structure of the Laplacian matrix} described in Section \ref{SectLapProp} and determined by the positive semidefiniteness of $\mathbf{L}$, the relation $\sum_{j=1}^N L_{ij}=0$ and the property that all the non-diagonal elements satisfied $L_{ij}\leq 0$. In order to maintain these properties, we require that the function $g(\mathbf{L})$ satisfies the following conditions:
\begin{itemize}
\item  {\bf Condition I:} The matrix $g(\mathbf{L})$ must be positive semidefinite, i.e., the eigenvalues of $g(\mathbf{L})$ are restricted to be positive or zero.
\item  {\bf Condition II:} The elements $g_{ij}(\mathbf{L})$, for $i,j=1,2,\ldots ,N$, should satisfy
\begin{equation}\label{gsumnull}
\sum_{j=1}^N g_{ij}(\mathbf{L})=0.
\end{equation}
\item {\bf Condition III:} All the non-diagonal elements of  $g(\mathbf{L})$ must satisfy $g_{ij}(\mathbf{L})\leq 0$.
\end{itemize}
The first condition is maintained if $g(x)\geq 0$ for $x\geq 0$, then $g(\mu_m)\geq 0$ for $m=1,2,\ldots, N$ and in this way the structure of (\ref{PositiveMu}) is also fulfilled for the eigenvalues of $g(\mathbf{L})$.
\\[2mm]
The second condition limits the function $g(\mathbf{L})$ to preserve the particular feature $\sum_{j=1}^N L_{ij}=0$ of the Laplacian matrix $\mathbf{L}$. As a direct consequence of this result, the Laplacian matrix has the eigenvector $|\Psi_1\rangle=\frac{1}{\sqrt{N}}
{\tiny\left[
\begin{array}{c}
 1 \\
 1 \\
\ldots \\
 1 \\
\end{array}
\right]}$
in Eq. (\ref{EigVectorL1}) associated to the smallest eigenvalue $\mu_1=0$. In addition,  the orthogonality condition $\langle \Psi_m|\Psi_1\rangle =0$ requires
\begin{equation*}
\sum_{j=1}^N \langle \Psi_m|j\rangle=0 \qquad \mathrm{for} \quad m=2,\ldots,N.
\end{equation*}
Then, by using Eq. (\ref{FunGLap}), we have
\begin{eqnarray*}
\sum_{j=1}^N g_{ij}(\mathbf{L})&=\sum_{j=1}^N\sum_{m=1}^N g(\mu_m)\langle i| \Psi_m\rangle \langle \Psi_m|j\rangle \\
&=\sum_{m=1}^N g(\mu_m)\langle i| \Psi_m\rangle \sum_{j=1}^N \langle \Psi_m|j\rangle,
\end{eqnarray*}
therefore
\begin{equation}\label{sumgL}
\sum_{j=1}^N g_{ij}(\mathbf{L})=g(\mu_1)\langle i| \Psi_1\rangle \sum_{j=1}^N\langle \Psi_1|j\rangle=g(\mu_1).
\end{equation}
Then, the condition II in Eq. (\ref{gsumnull}) is fulfilled if the function $g(x)$ satisfies $g(0)=0$.
\\[2mm]
Until now, we have determined the first two conditions for the function $g(\mathbf{L})$. However, these conditions are not at all sufficient to guarantee
that $g(\mathbf{L})$ remains with all non-diagonal elements satisfying $g_{ij}(\mathbf{L})\leq 0$ as requires the condition III.
\subsection{Non-negative symmetric matrices}
In this part we explore the necessary conditions for an admissible class of functions $g(\mathbf{L})$ maintaining the condition III: $g_{ij}(\mathbf{L})\leq 0$, for $i\neq j$. Then, let us consider the matrix $\mathbf{B}(t)$ given by
\begin{equation}\label{MatrixB}
\mathbf{B}(t)=\kappa\mathbb{I}-t\mathbf{L},
\end{equation}
here $t$ is a real value in the interval $0\leq t \leq 1$ and $\kappa$ is a parameter that satisfies the condition $k_{\mathrm{max}}<\mu_N\leq\kappa$. The lower limit in the last inequality is determined by the condition in Eq. (\ref{mumax}) for the largest eigenvalue $\mu_N$.
\\[2mm]
In the definition in Eq. (\ref{MatrixB}), we observe that all the elements of $\mathbf{B}(t)$ satisfy $B_{ij}(t)\geq 0$ for $t$ in the interval $0\leq t \leq 1$; also, this condition is maintained for all the integer powers of  $\mathbf{B}(t)$, i.e. $(\mathbf{B}^n)_{ij}(t)\geq 0$ for $n=1,2,3\ldots$. On the other hand, from the spectral decomposition of the matrix $\mathbf{B}(t)$ in Eq. (\ref{MatrixB}), we observe that
\begin{equation}
\mathbf{B}(t)=\sum_{m=1}^N (\kappa-t \mu_m)|\Psi_m\rangle\langle\Psi_m|
\end{equation}
is positive definite with eigenvalues $\kappa-t\mu_m>0$ for $m=1,2,\ldots ,N$ and $0\leq t\leq 1$, in this way positive definiteness  is also preserved for all the integer powers of  $\mathbf{B}(t)$.
\\[2mm]
Let now be $h(x)=\sum_{n=0}^\infty \frac{h^{(n)}(0)}{n!} x^n >0$ a positive scalar scalar function with all non-negative derivatives $h^{(n)}(x)=\frac{d^n}{dx^n} h(x)\geq 0$  ($n=0,1,2,\ldots$) for $x \geq 0 $ especially on the interval of the spectrum of eigenvalues of $\mathbf{L}$. For this function we have
\begin{equation}\label{hmatrixBseries}
h(\mathbf{B}(t))=h(\kappa\mathbb{I}-t\mathbf{L})=\sum_{n=0}^\infty \frac{h^{(n)}(0)}{n!}(\kappa\mathbb{I}-t\mathbf{L})^n,
\end{equation}
where we observe that each term in the series has only non-negative matrix elements 
\begin{equation}
\left[\frac{h^{(n)}(0)}{n!}(\kappa\mathbb{I}-t\mathbf{L})^n\right]_{ij}\geq 0 \qquad \mathrm{for} \quad i,j=1,\ldots ,N.
\end{equation}
Hence, by using the Eq. (\ref{hmatrixBseries}), the matrix $h(\mathbf{B}(t))$ has all non-negative elements
\begin{equation}\label{hij_nonnegative}
h_{ij}(\kappa\mathbb{I}-t\mathbf{L})\geq 0 \qquad \mathrm{for} \quad i,j=1,\ldots N \quad \mathrm{and} \quad 0\leq t\leq 1
\end{equation}
and also maintains the positive definiteness property with eigenvalues $h(\kappa-t\mu_m)\geq 0$ for $m=1,\ldots, N$.
\\[2mm]
Now, we consider the function $H(x)=H(0)+\sum_{n=0}^\infty \frac{h^{(n)}(0)}{n+1!}x^{n+1}$ defined as a primitive of the function $h(x)$ and, when we choose $H(0)=0$ we have $H(x)=\int_0^x h(t)dt$. Following the same reasoning that leads to the result in Eq. (\ref{hij_nonnegative}), we see that all the matrix elements of $H(\mathbf{B}(t))$ are also non-negative
\begin{equation}\label{Hij_nonnegative}
H_{ij}(\kappa\mathbb{I}-t\mathbf{L})\geq 0 \qquad \mathrm{for} \quad i,j=1,\ldots, N \quad \mathrm{and} \quad 0\leq t\leq 1
\end{equation}
and again $H(\mathbf{B}(t))$ is a positive definite matrix.
\\[2mm]
Once introduced the functions $h(x)$ and $H(x)$, we can establish one of our main results with respect to functions that preserve the structure of the Laplacian matrix. First, let us consider the integral
\begin{eqnarray}
g(x\mathbf{L})&=\int_0^x h(\kappa\mathbb{I}-t\mathbf{L})\mathbf{L}dt=-H(\kappa\mathbb{I}-t\mathbf{L})\bigg|_0^x  \\ \label{gxintegral}
&=H(\kappa\mathbb{I})-H(\kappa\mathbb{I}-x\mathbf{L})
\qquad \mathrm{for} \quad 0\leq x\leq 1.
\end{eqnarray}
From our previous analysis and the result in Eq. (\ref{gxintegral}), we notice that $h(\kappa\mathbb{I}-t\mathbf{L})\mathbf{L}$ and hence $g(x\mathbf{L})$ are positive semidefinite matrices, having the eigenvalues $h(\kappa-\mu_m)\mu_m$, and $H(\kappa)-H(\kappa-x\mu_m)$, respectively, and maintaining the eigenvalue $0$ for $m=1$ in the matrix  $g(x\mathbf{L})$. In this way, the function in Eq. (\ref{gxintegral})  fulfills conditions I and II.
\\[2mm]
Now, we introduce the function $f(z)=h(\kappa-z)$  for which we have the matrix function $f(\mathbf{L})=h(\kappa \mathbb{I}-\mathbf{L})$ and, by using Eq. (\ref{gxintegral}), we have
\begin{equation}\label{g_integral}
g(x)=\int_0^x f(z)dz=\int_0^x h(\kappa-z)dz=H(\kappa)-H(\kappa-x).
\end{equation}
According to the construction of the function $h(x)$, $f(z)$ satisfies
\begin{equation}\label{fmono}
(-1)^n \frac{d^{n}}{dz^n}f(z)=\frac{d^{n}}{dz^n}h(x-z)\geq 0 , \hspace{0.5cm} n=1,2,..
\end{equation}
and non-vanishing $h(x-z) > 0$ for the order $n=0$ which guarantees that all coefficients in Eq. (\ref{hmatrixBseries}) are strictly non-negative.
On the other hand, from Eq. (\ref{gxintegral}), we see that the first term in this relation $H_{ij}(\kappa\mathbb{I})=\delta_{ij}H(\kappa)$ is diagonal. Then, as a consequence of Eq. (\ref{Hij_nonnegative}), the off-diagonal elements in Eq. (\ref{gxintegral}) satisfy
\begin{equation}
g_{ij}(x\mathbf{L})=[H(\kappa\mathbb{I})-H(\kappa\mathbb{I}-x\mathbf{L})]_{ij}=-H_{ij}(\kappa\mathbb{I}-x\mathbf{L})\leq 0 
\end{equation}
for $i\neq j$ and $0\leq x\leq 1$. Hence, the matrix function $g(\mathbf{L})$ (in this particular case $x=1$) fulfills additionally to conditions I and II, also the restriction to preserve the Laplacian structure given by condition III. The eigenvalues of $g(\mathbf{L})$  are positive because the monotony of $H(x)$, then $H(\kappa)-H(\kappa-\mu_m)>0$ for $m=2,\ldots, N$ and is null for $m=1$ since $\mu_1=0$. 
\subsection{Completely monotonic functions}
From the results established above, we know that the conditions I, II, III that satisfy the Laplacian matrix are maintained for matrix functions $g(\mathbf{L})$ determined by
\begin{equation}
g(\mathbf{L})=H(\kappa \mathbb{I})-H(\kappa \mathbb{I}-\mathbf{L}).
\end{equation}
The function $g(x)$ can be expressed in terms of the function $f(x)$ that fulfills the following necessary and sufficient conditions
\begin{equation} \label{choice1}
\frac{d}{dx}g(x) = f(x) > 0 ,\qquad \mu_1=0 ,\quad 0\leq x \leq \mu_N < \kappa
\end{equation}
and further, from Eq. (\ref{fmono}) we have 
\begin{equation} \label{choice2}
 (-1)^n\frac{d^n}{dx^n}f(x) \geq 0 ,\qquad 0\leq x \leq \mu_N < \kappa
\end{equation}
for $n=1,2,..$, i.e. $\frac{d^n}{dx^n}f(x)$ has alternating sign if non-zero and where (\ref{choice1}) is non-vanishing especially on the entire spectral interval $0\leq x \leq \mu_N$ of $\mathbf{L}$. In addition, due to Eq. (\ref{g_integral}), the matrix functions $g(x)$ are then given by the following integral
\begin{equation}\label{choice3}
g(x) =\int_0^xf(z)\, dz
\end{equation}
which is the primitive of $f(x)$ with $g(0)=0$ and with $g(x) >0$ for $x>0$. This last condition preserves the eigenvalue zero of $\mathbf{L}$ in the matrix function $g(\mathbf{L})$. On the other hand, the positiveness of $f(x)$ guarantees the positiveness of the remaining $N-1$ eigenvalues, and Eq. (\ref{choice3}) together with the relation in Eq. (\ref{choice1}) maintain conditions I to III. 
\\[2mm]
The results in Eqs. (\ref{choice1}) and (\ref{choice2}) indicate that $f(x)$ is a {\it monotonously decreasing} function with $x$ over the interval
of its definition $0\leq x\leq \kappa$, {\it and remains positive} over the spectral interval of eigenvalues with $0\leq \mu_m \leq \mu_N < \kappa$ and can hence be written as $f(x)=h(-x)$ where $f(-x)=h(+x)$ is a {\it monotonously increasing} function with $x$
where $\frac{d^n}{dx^n}f(-x) \geq 0$ as consequence of Eq. (\ref{choice2}). The condition of positiveness of $f(x)$, namely (\ref{choice1}) implicitly accounts for the parameter $\kappa > \mu_N$. On the other hand, the function $g(x)$ in Eq. (\ref{choice3}) has the following general structure
\begin{equation}\label{general}
g(x)= \int_0^xh(-t){\rm d}t = H(0)-H(-x) \geq 0
\end{equation}
and, in this way, admissible matrix functions $g(\mathbf{L})$ fulfilling I-III are given by
\begin{equation}\label{general}
g(\mathbf{L})= H(0){\mathbb{I}}-H(-\mathbf{L}) = \sum_{m=1}^N \left(H(0)-H(-\mu_m)\right)\left|\Psi_m\rangle\langle\Psi_m \right| .
\end{equation}
Condition III is satisfied due to the fact that $H(-\mathbf{L})$ is a non-negative matrix. We see further in Eq. (\ref{general}) that the eigenvalue for $m=1$ is vanishing whereas those for $m=2,3,\ldots,N$ remain positive as $H(0)-H(-x) >0$ for $x>0$, i.e. conditions I-III are fulfilled by Eq. (\ref{general}). Functions of the Laplacian maintaining these conditions have been little explored, Michelitsch et. al. have analyzed conditions I and II in connection with the non-locality generated by matrix functions in lattices \cite{Michelitsch2014Nonlocal}. On the other hand, Micchelli and Willoughby in \cite{Micchelli1979}  gave the conditions on a function $f$ so that if the matrix $\mathbf{M}$ is symmetric and nonnegative so is $f(\mathbf{M})$. Hence, through the Eq. (\ref{MatrixB}) established between the Laplacian and the non-negative matrix $\mathbf{B}$, the functions $g(\mathbf{L})$ of the Laplacian matrix preserve the conditions I-III.
\\[2mm]
%
%
A function $f(x)$ defined on $x \geq 0$ is said to be {\it completely monotonic} if it has derivatives $f^{(n)}(x)$ for $n=0, 1,2, \ldots$ and $(-1)^n f^{(n)}(x)\geq 0$ for all $x>0$. We also mention that functions having a completely monotonic derivative are referred to as {\it Bernstein functions} \cite{bernstein1929}.
Good Laplacian functions $g(x)$ are the class of Bernstein functions with $g(x=0)=0$ and strictly non-vanishing first derivative $\frac{d}{dx}g(x)=f(x) > 0$ for all $x\geq 0$. It follows that not all Bernstein functions are always good Laplacian functions, but good Laplacian functions $g(x)$ are always Bernstein functions: Generally Bernstein functions are allowed to be non-vanishing at $x=0$ and to have pointwise vanishing first derivatives for instance at $x=0$, whereas these two properties are forbidden for good Laplacian functions.
\\[2mm]
There exists different types of completely monotonic functions that in combination with the integral in Eq. (\ref{choice3}) allow to define functions $g(\mathbf{L})$ that maintain the Laplacian structure. In the following sections we explore $g(\mathbf{L})$ in connection with random walk strategies, we analyze the particular cases
\begin{itemize}
\item The function $f(x)=(\beta+1)x^\beta$ with $\beta\leq 0$ fulfills the condition in Eq. (\ref{fmono}) for a completely monotonic function. Consequently, the integral in Eq. (\ref{choice3}) allows to obtain $g(x)=x^{\beta+1}$; however, the additional condition $g(x)>0$ for $x>0$ requires $-1\leq\beta\leq 0$. Therefore, the function
\begin{equation}\label{g_power}
g(x)=x^\gamma \qquad \mathrm{for} \quad 0<\gamma \leq 1,
\end{equation}
maintains the structure of the Laplacian described in conditions I-III. In the following section we will study this function in connection with the fractional Laplacian of a graph \cite{RiascosMateosFD2014,RiascosMateosFD2015,Michelitsch2017PhysA,Michelitsch2017PhysARecurrence}.

\item For the completely monotonic function $f(x)=\frac{\alpha}{1+\alpha x}$ with $\alpha>0$, by using the integral in Eq. (\ref{choice3}), we have
\begin{equation}\label{loggfunction}
g(x)=\log(1+\alpha x)             \qquad \mathrm{with} \quad \alpha>0.
\end{equation}

\item Another completely monotonic function is determined by the exponential $f(x)=a e^{-ax}$ with $a>0$ for which we see that Eq. (\ref{fmono}) is satisfied. The corresponding function $g(x)$ that preserves the Laplacian structure is
\begin{equation}\label{expgfunction}
g(x)=1-e^{-ax}  \qquad \mathrm{with} \quad a>0.
\end{equation}
\end{itemize}
The following observation with respect to the admissible functions $g(\mathbf{L})$ appears noteworthy. Let us briefly consider matrix functions defined by powers of $\mathbf{L}$, namely $g_{\beta}(\mathbf{L}) = \mathbf{L}^{\beta}$. As we saw above only power functions with exponents $0<\beta \leq 1$ are admissible. Powers with $\beta >1$ are not  since they do not fulfill Eq. (\ref{choice2}). From this observation follows that admissible functions $g(x)$ obey (up to positive multiplyers) for small arguments 
\begin{equation}
\label{case1}
g(x) \sim x^{\gamma} ,\hspace{1cm}x\rightarrow 0+, \hspace{1cm} 0<\gamma \leq 1 .
\end{equation}
The lowest order in the expansion of an admissible function $g(x)$ either starts with $x$ (type (i)) i.e. integer $\gamma=1$, or with $x^{\gamma}$ ($0<\gamma<1$, type (ii)) when it is non-integer (fractional). The expansion of $g(x)$ (up to unimportant positive multipliers) for type (i) functions are of the form $g(x)= x+{\tilde g}(x)..$, whereas type (ii) functions have expansions that write as $g(x)= x^{\gamma} +{\tilde g}(x).. $ ($0<\gamma<1$). The parts ${\tilde g}(x)$ contain only powers greater than 1 in case (i), and greater than $\gamma$ in case (ii), respectively. The classes (i) and (ii) are the only two classes of functions that are admissible.  In view of above considered examples, the functions $1-e^{-x}$ and $\log(1+\alpha x)$ are type (i) functions, whereas $x^{\gamma}$ ($0<\gamma<1$) is of type (ii).
\\[2mm]
In appendix C we demonstrate that the lowest power in the expansion of $g(\mathbf{L})$ determines the dominant asymptotic transition probability for long-range steps on sufficiently large networks $N\rightarrow\infty$. 
We further show there that functions $g(\mathbf{L})=\mathbf{L}+{\tilde g}(\mathbf{L})$  of type (i) contain an internal length-scale defined by the local information of Laplacian $\mathbf{L}$. This type of non-locality depends on that length-scale and by increasing the size of the network, the Laplacian functions of type (i) become quasi-local and the type (i) random walk strategy becomes similar to a normal random walk with emerging {\it Brownian motions} (normal diffusion) in the limit of large networks $N\rightarrow\infty$.
\\[2mm]
In contrast functions of type (ii) define a fractional type of non-locality with $g(\mathbf{L})=\mathbf{L}^{\gamma}+{\tilde g}(\mathbf{L})$ ($0<\gamma<1$) which becomes asymptotically scale-free (asymptotically self-similar) in the limit of large networks $N\rightarrow\infty$: The asymptotic scale-freeness wipes out in the limit of infinite networks any local information on $\mathbf{L}$ and in this sense is universal. We will see in appendix C that type (ii) non-locality leads to asymptotic emergence of L\'evy flights (anomalous diffusion) on large networks $N\rightarrow\infty$. The type (ii) non-locality due to its asymptotic scale-freeness cannot be `localized' as in case (i) by increasing the size $N$ of the network.  The type (ii) non-locality thus remains `stable' when increasing the size of the network. We conjecture that only type (ii) non-locality can maintain communication in dynamically growing complex networks such as living structures and time-evolving networks whereas under type (i) non-locality 
far distant nodes become disconnected.
\\[2mm]
Again we emphasize that only these two classes of functions $g(\mathbf{L})$ type (i) and type (ii) constitute good functions to define random walks. 
For the asymptotic behavior of the walk emerging in the limit of an inifinite network only the lowest orders are relevant, i.e. $\mathbf{L}$ for type (i), and $\mathbf{L}^{\gamma}$ ($0<\gamma<1$) for type (ii) functions, respectively. The part ${\tilde g}(\mathbf{L})$ containing the higher orders in $\mathbf{L}$ becomes irrelevant in the infinite network limit.

We hence refer functions $g$ of type (i) to as Brownian type functions, and functions of type (ii) to as L\'evy type functions. These two classes of admissible functions have their counterparts in Gaussian (type (i)) and L\'evy-stable (type (ii)) distributions. Further properties and analysis of this issue is given in appendix C. 
\\[2mm]
The formalism introduced is general and can be applied to completely monotonic functions that once integrated to obtain $g(x)$ through Eq. (\ref{choice3}) are well defined and satisfy $g(0)=0$. Other examples of completely monotonic functions are the modified Bessel function of the first kind, the Mittag Leffler function that appears in the context of fractional calculus, among many others  \cite{Miller2001}. In addition, the composition of  completely monotonic functions produces other types of functions that fulfill the condition in Eq. (\ref{choice2}), see details in \cite{Miller2001,Merkle2014}.
\subsection{General properties of $g(\mathbf{L})$}
Once identified functions that maintain the structure of the Laplacian matrix $\mathbf{L}$, in this part we discuss some general properties of $g(\mathbf{L})$.
\subsubsection{Diagonal elements (generalized degree).}
By definition, diagonal elements of the matrix $g(\mathbf{L})$ are positive and, in analogy with the Laplacian matrix $\mathbf{L}$, we denote the diagonal elements of $g(\mathbf{L})$  as the {\it generalized degree} associated to the function $g$ as
\begin{equation*}
\mathcal{K}_i\equiv g_{ii}(\mathbf{L}).
\end{equation*}
Now, as a direct consequence of Eq. (\ref{sumgL}) and the condition $g(0)=0$, we have
\begin{equation}
0=\sum_{j=1}^N g_{ij}(\mathbf{L})=\mathcal{K}_i+\sum_{j\neq i}g_{ij}(\mathbf{L})
\end{equation}
with $i=1,2,\ldots, N$. Therefore, the  generalized degree $\mathcal{K}_i$ takes the form
\begin{equation}\label{KdegreeSumg}
\mathcal{K}_i=-\sum_{j\neq i}g_{ij}(\mathbf{L}).
\end{equation}
On the other hand, the average of the generalized degree satisfies
\begin{equation}\label{AverageK}
\langle \mathcal{K}\rangle=\frac{1}{N}\sum_{i=1}^N \mathcal{K}_i=\frac{1}{N}\Tr(g(\mathbf{L}))
=\frac{1}{N}\sum_{i=1}^N g(\mu_i),
\end{equation}
showing that $\langle \mathcal{K} \rangle$ can be calculated directly from the spectrum of the Laplacian matrix $\mathbf{L}$. In the general case, the degree $\mathcal{K}_i$ is a quantity that not only incorporates information on the nearest neighbors of $i$, but also includes information of the whole structure. This non-locality  is explored in the following part.
\subsubsection{Functions $g(\mathbf{L})$ for regular graphs.}
Now, in order to understand the structure of the matrix $g(\mathbf{L})$ we analyze the particular case of regular networks. For this type of structures, the degree $k$ (number of connections that a node has)  is a constant and the Laplacian matrix $\mathbf{L}$ takes the form
\begin{equation}
\mathbf{L}=k\mathbb{I}-\mathbf{A}.
\end{equation}
Furthermore, the series expansion of $g(x)$ is given by
\begin{equation}\label{gseries}
g(x)=\sum_{l=1}^\infty c_l x^l,
\end{equation}
where the constants $c_l$ for $l=1,2,\ldots$, are particular for each function $g(x)$ and especially $c_0=0$ reflecting $g(0)=0$. Now, in terms of the series expansion in Eq. (\ref{gseries}), we obtain the following result for regular networks
\begin{eqnarray}\nonumber
g(\mathbf{L})&=\sum_{l=1}^\infty c_l (k\mathbb{I}-\mathbf{A})^l
=\sum_{l=1}^\infty c_l\sum_{m=0}^l {l \choose m}(k\mathbf{I})^{l-m}(-1)^m\mathbf{A}^m\\
\label{SumApowers}
&=\sum_{l=1}^\infty \sum_{m=0}^l c_l  {l \choose m}k^{l-m}(-1)^m\mathbf{A}^m.
\end{eqnarray}
The Eq. (\ref{SumApowers}) establishes a connection between the matrix $g(\mathbf{L})$ with the integer powers of the adjacency matrix $\mathbf{A}^m$ for $m=1, 2,\ldots$ for which the element $(\mathbf{A}^m)_{ij}$  is the number of all the possible trajectories connecting the nodes $i$, $j$ with $m$ links \cite{GodsilBook}. On the other hand, the diagonal element $(\mathbf{A}^m)_{ii}$  is the number of closed trajectories with $m$ links on the network that start in the node $i$ and end in the same node \cite{GodsilBook}. In this way, Eq. (\ref{SumApowers}) reveals how the functions $g(\mathbf{L})$ changes  the local character of the Laplacian matrix $\mathbf{L}$ to a long-range operator. The resulting matrix is appropriate to define a diversity of dynamical processes with non-local interactions on networks. As particular cases of the application of this formalism we have the fractional diffusion and the quantum transport  on networks \cite{RiascosMateosFD2014,RiascosMateos2015}, the diffusion on finite and 
infinite lattices \cite{Michelitsch2016,Michelitsch2016Chaos,Michelitsch2017PhysA,Michelitsch2017PhysARecurrence}, and semi-supervised learning algorithms \cite{SdeNigris2017}.
\section{Random walk strategies and $g(\mathbf{L})$}
\label{RW}
In this section we study discrete time random walks with a transition probability $\pi_{i\to j}$ that the walker moves from node $i$ to node $j$ defined in terms of the matrix $g(\mathbf{L})$. We start with the discrete time master equation that describes the dynamics of a Markovian random walker on a network \cite{HughesBook}
\begin{equation}\label{master}
P_{ij}(t+1) = \sum_{l=1}^N  P_{il} (t) \pi_{l\to j} \ ,
\end{equation}
where $P_{ij}(t)$ is the occupation probability to find the random walker in $j$ at time $t$ starting from $i$ at $t=0$. The time $t$ is restricted to integer values denoting the number of steps made by the random walker. All the statistical information of how the random walker moves in the network is contained in the transition matrix $\bf{\Pi}$ with elements $\pi_{i\to j}$. In the following part, we introduce a general random walker that defines $\pi_{i\to j}$ with functions of the Laplacian matrix of a network.
\subsection{Transition probability matrix}
We explore a random walker that moves on a simple connected network with $N$ nodes described by the adjacency  matrix $\mathbf{A}$. At each step, a walker moves randomly from a node $i$ to a site $j$ following a strategy defined by the transition probability
\begin{eqnarray}
\pi_{i \to j}
&=
\frac{1}{\sum_{l\neq i}g_{il}(\mathbf{L})}
\left\{
\begin{array}{ll}
0                                                                        &\mathrm{for}  \quad i=j,\\
g_{ij}(\mathbf{L})  \qquad      &\mathrm{for}   \quad i\neq j.
\end{array}\right .
\end{eqnarray}
Here $\frac{1}{\sum_{l\neq j}g_{il}(\mathbf{L})}$ is a normalization factor that guaranties that the probability to hop from node $i$ to any site of the network is $1$. On the other hand, the particular case  $\pi_{i \to i}=0$ establishes that the random walker changes its position at each step. Now, by using the result in Eq. (\ref{KdegreeSumg}) for the generalized degree $\mathcal{K}_i=-\sum_{l\neq i}g_{il}(\mathbf{L})$, we have for the transition probability $\pi_{i \to j}$
\begin{equation}\label{wij_g}
\pi_{i \to j}=\delta_{ij}-\frac{g_{ij}(\mathbf{L})}{\mathcal{K}_i}.
\end{equation}
The conditions described in Section \ref{FunctionSection} allow to define properly the transition probabilities in Eq. (\ref{wij_g}). For example, by using Condition III, we know that non-diagonal elements of $g(\mathbf{L})$ are negative or null and in this way $\pi_{i \to j}$ in Eq. (\ref{wij_g}) always can be interpreted as a non-negative transition probability.
\\[3mm]
%
%
Once defined a general strategy for random walks on networks, the study of particular cases helps us to understand the non-local character of the random walk strategies that emerge from different choices of $g(\mathbf{L})$ in Eq. (\ref{wij_g}).
\subsubsection{Fractional Laplacian.}  
\begin{figure}[ht!]
\begin{center}
\includegraphics[width=0.9\textwidth]{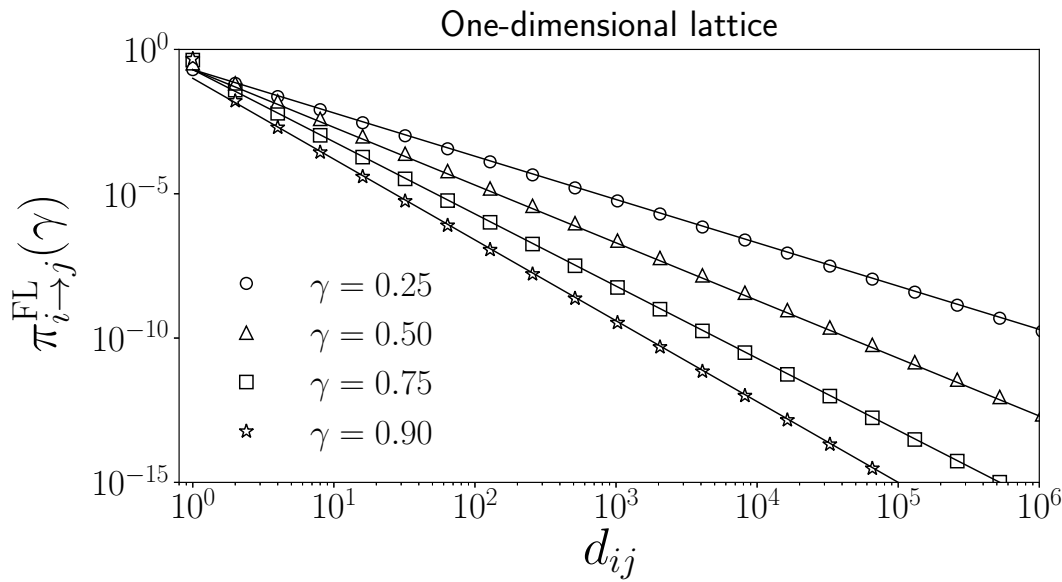}
\end{center}
\vspace{-6mm}
\caption{\label{Fig1} Transition probability $\pi^{\mathrm{FL}}_{i\to j}(\gamma)$ as a function of the distance $d_{ij}$ for the random walk strategy generated by using the fractional Laplacian $g(\mathbf{L})=\mathbf{L}^\gamma$ for an infinite ring (1D lattice with periodic boundary conditions). Continuous lines represent the inverse power-law relation $\pi^{\mathrm{FL}}_{i\to j}(\gamma)\propto d_{ij}^{-1-2\gamma}$.}
\end{figure} 
In this case, a random walk strategy is defined in terms of the  function $g(x)=x^\gamma$ with $0<\gamma\leq 1$ discussed in Eq. (\ref{g_power}). Through this function is obtained  the {\it Fractional Laplacian of a graph} \cite{RiascosMateosFD2014}
\begin{equation}\label{Lgamma_mat}
g(\mathbf{L})=\mathbf{L}^{\gamma}, 
\end{equation}
where $\gamma$ is a real number that satisfies $0<\gamma<1$. The fractional Laplacian matrix defined in this index range way is the relevant part and 
proto-example for a type (ii) Laplacian function introduced in Section \ref{FunctionSection}.
The matrix in Eq. (\ref{Lgamma_mat}) has been studied in the context of fractional diffusion on general networks and lattices \cite{RiascosMateosFD2014,RiascosMateosFD2015,Michelitsch2017PhysA,Michelitsch2017PhysARecurrence}. The resulting random walk is defined by a transition probability $\pi^{\mathrm{FL}}_{i\to j}(\gamma)$ given by
\begin{equation}\label{wijfrac}
\pi^{\mathrm{FL}}_{i\to j}(\gamma)=\delta_{ij}-\frac{(\mathbf{L}^\gamma)_{ij}}{k_i^{(\gamma)}}\qquad 0<\gamma\leq 1.
\end{equation}
There is an important limiting case: In Eq. (\ref{wijfrac}), when we have the limit $\gamma \to 1$, the transition probability $\pi^{\mathrm{NRW}}_{i\to j}=\delta_{ij}-\frac{L_{ij}}{k_i}=\frac{A_{ij}}{k_i}$ which corresponds to the normal random walk (NRW) on networks, previously studied by other authors \cite{NohRieger}, describing local transitions only to nearest neighbors with equal probability, that is, inversely proportional to the degree $k_i$ of the node $i$. In the relation in Eq. (\ref{wijfrac}), the fractional Laplacian $\mathbf{L}^\gamma$ is calculated by using eigenvalues and eigenvectors of $\mathbf{L}$ in Eq. (\ref{FunGLap}). On the other hand, the diagonal elements of $\mathbf{L}^\gamma$ constitute a particular type of generalized degree.  In this way, the {\it fractional degree} $k_i^{(\gamma)}$ of the node $i$ is \cite{RiascosMateosFD2014}
\begin{equation}\label{FracDegreeGeneral}
 k_i^{(\gamma)}\equiv(\mathbf{L}^{\gamma})_{ii}=\sum_{m=2}^N \mu_m^{\gamma}\langle i\left |\Psi_m\right\rangle\left\langle \Psi_m\right| i\rangle.
\end{equation}
In order to see the type of transition probabilities that emerge from the definition (\ref{wijfrac}), in Figure \ref{Fig1} we calculate the values of $\pi^{\mathrm{FL}}_{i\to j}(\gamma)$ for an infinite one-dimensional lattice with periodic boundary conditions. We describe this particular case in appendix A for which we can explore the fractional Laplacian analytically due to the fact that the eigenvalues and eigenvectors for this case are known. Our results in  Figure \ref{Fig1} reveal the relation $\pi^{\mathrm{FL}}_{i\to j}(\gamma) \sim d_{ij}^{-1-2\gamma}$, for the cases explored, where the distance $d_{ij}$ is the length of the shortest path connecting the nodes $i$ and $j$, this relation is valid for distances $d_{ij}\gg 1$. 
\\[2mm]
In the general case, the fractional random walk is the process associated to the fractional diffusion on networks and the transition probabilities in Eq. (\ref{wijfrac}) define a navigation strategy with long-range displacements on the network \cite{RiascosMateosFD2014}. The case of infinite $n$-dimensional lattices with periodic boundary conditions has been addressed in different in contexts \cite{Michelitsch2016,Michelitsch2016Chaos,Michelitsch2017PhysA,Michelitsch2017PhysARecurrence}.  In this case, we have the analytical relation \cite{Michelitsch2017PhysA}
\begin{equation}\label{LevyLattices}
\pi^{\mathrm{FL}}_{i\to j}(\gamma)\sim d_{ij}^{-n-2\gamma}\qquad \mathrm{for} \quad d_{ij}\gg 1.
\end{equation}
The result in Eq. (\ref{LevyLattices}) establishes a connection between L\'evy flights on networks \cite{RiascosMateosLF2012} and the fractional strategy defined by Eq. (\ref{wijfrac}). A detailed analysis of the fractional Laplacian of graphs and its relation with long-range navigation on networks and applications is presented in references \cite{RiascosMateosFD2014,RiascosMateosFD2015,Michelitsch2016,Michelitsch2016Chaos,Michelitsch2017PhysA,deNigris2016,DeNigris2017,Michelitsch2017PhysARecurrence}.

\subsubsection{Logarithmic functions of the Laplacian.} In this part we explore the resulting dynamics for the function $g(x)=\log(1+\alpha x)$ presented in Eq. (\ref{loggfunction}) that fulfills with all the conditions described in Section \ref{FunctionSection}. In this case we have
\begin{figure}[b!]
\begin{center}
\includegraphics[width=0.9\textwidth]{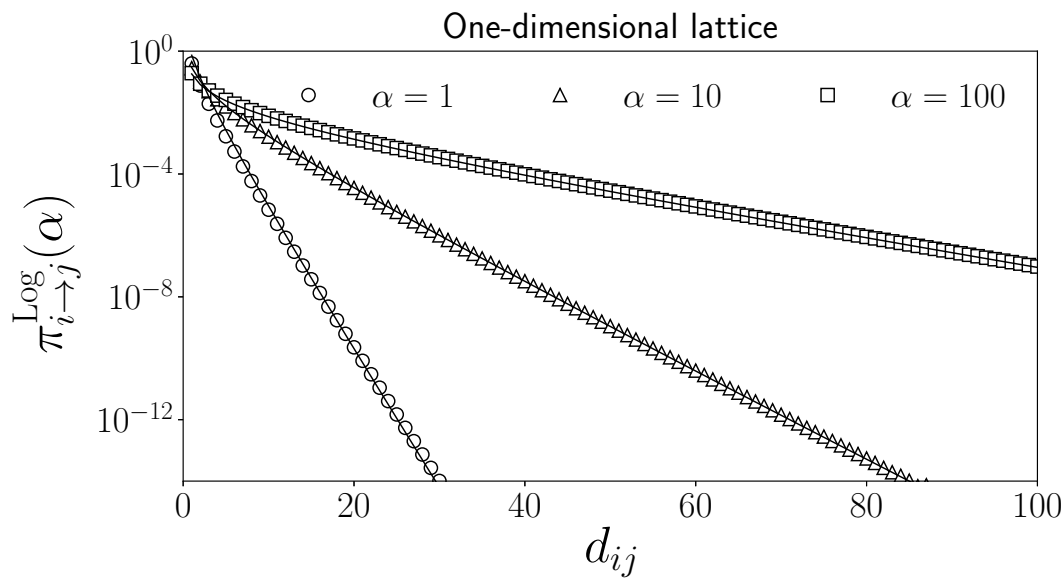}
\end{center}
\vspace{-6mm}
\caption{\label{Fig2} Transition probability $\pi^{\mathrm{Log}}_{i\to j}(\alpha)$ as a function of the distance $d_{ij}$ for the random walk strategy defined in terms of the logarithmic function $g(\mathbf{L})=\log\left(\mathbb{I}+\alpha \mathbf{L}\right)$ for an infinite ring (1D lattice with periodic boundary conditions). Continuous lines represent the asymptotic result $\pi^{\mathrm{Log}}_{i\to j}(\alpha)\propto \frac{e^{-d_{ij}/\sqrt{\alpha}}}{d_{ij}}$.}
\end{figure} 
\begin{equation}\label{gLog}
g(\mathbf{L})=\log\left(\mathbb{I}+\alpha \mathbf{L}\right) \qquad \mathrm{for} \quad \alpha>0
\end{equation}
and the resulting random walk strategy is given by
\begin{equation}\label{LogLattices}
\pi^{\mathrm{Log}}_{i\to j}(\alpha)=\delta_{ij}-\frac{\log\left(\mathbb{I}+\alpha \mathbf{L}\right)_{ij}}{\log\left(\mathbb{I}+\alpha \mathbf{L}\right)_{ii}}.
\end{equation}
By using the methods described in appendix A, we calculate analytically the transition probabilities $\pi^{\mathrm{Log}}_{i\to j}(\alpha)$ for an infinite one-dimensional lattice with periodic  boundary conditions. In Figure \ref{Fig2} we depict the results obtained by numerical integration of the expression in Eq. (\ref{Lijintegral}) and the definition in Eq. (\ref{wij_g}). In Figure \ref{Fig2},   continuous lines represent the approximation $\pi^{\mathrm{Log}}_{i\to j}(\alpha)\propto \frac{e^{-d_{ij}/\sqrt{\alpha}}}{d_{ij}}$. It is worth to notice that this relation takes the form   $\pi^{\mathrm{Log}}_{i\to j}(\alpha)\propto e^{-d_{ij}/\sqrt{\alpha}}$ for large displacements on the lattice $d_{ij}\gg 1$. A random walk strategy including transition probabilities with a similar exponential relation is introduced and explored in detail for different types of graphs by Estrada et. al. in \cite{Estrada2017}.

\subsubsection{Exponential functions of the Laplacian.} Now we introduce a random walker defined in terms of the function in Eq. (\ref{expgfunction}). We explore the function $g(x)=1-e^{-a x}$ defined in terms of an exponential that allows to define a random walk strategy by using Eq. (\ref{wij_g}). In this case we have the matrical function
\begin{equation}\label{gExp}
g(\mathbf{L})=\mathbb{I}-e^{-a\mathbf{L}} \qquad \mathrm{for} \quad a>0
\end{equation}
and the corresponding transition probabilities $\pi^{\mathrm{Exp}}_{i\to j}(a)$ determined by
\begin{equation}\label{ExpWij}
\pi^{\mathrm{Exp}}_{i\to j}(a)=\delta_{ij}-\frac{\left(\mathbb{I}-e^{-a\mathbf{L}}\right)_{ij}}{\left(\mathbb{I}-e^{-a\mathbf{L}}\right)_{ii}} \qquad \mathrm{with} \quad a> 0.
\end{equation}
\begin{figure}[t!]
\begin{center}
\includegraphics[width=0.9\textwidth]{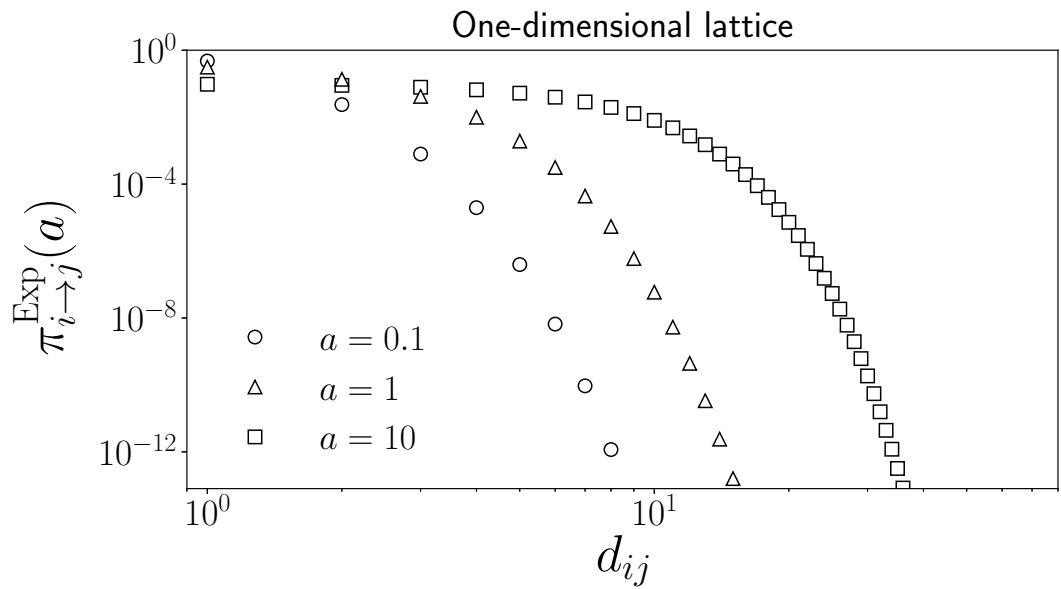}
\end{center}
\vspace{-6mm}
\caption{\label{Fig3} Transition probability $\pi^{\mathrm{Exp}}_{i\to j}(a)$ as a function of the distance $d_{ij}$ for the random walk defined in terms of the function $g(\mathbf{L})=\mathbb{I}-e^{-a\mathbf{L}}$ for a 1D lattice with periodic boundary conditions.}
\end{figure} 
In order to see the non-local behavior of the transition probabilities in Eq. (\ref{ExpWij}),  in Figure \ref{Fig3} we depict the transition probabilities $\pi^{\mathrm{Exp}}_{i\to j}(a)$ for an infinite one-dimensional lattice (see appendix A for details). In this case, the results reveal long-range transitions on the lattice and how the parameter $a$ controls the dynamics ranging from a nearly local case, in the limit $a\to 0$, to long-range displacements when $a\gg 1$. 
\\[2mm]
Finally, it is worth to mention that the exponential of the Laplacian matrix $e^{-a \bf{L}}$ is common in the study of classical transport on lattices and networks and is also called the heat kernel because of its interpretation as a diffusion process related to the heat equation  \cite{FoussBook2016}.  In addition, draws the attention that functions like the exponential $e^{-a \bf{L}}$ and the regularized Laplacian $(\mathbb{I}+\alpha \mathbf{L})^{-1}$ are used as kernels to compute similarities between nodes of an undirected graph \cite{FoussBook2016}. In the formalism discussed in this section these two type of functions lead respectively, through integration, to the exponential and logarithmic transition probabilities discussed in Eqs. (\ref{ExpWij}) and (\ref{LogLattices}).
\section{Global characterization}
Once described the properties of functions $g(\mathbf{L}) $ that allow to define random walk strategies and explored some particular cases; in this section, we characterize the capacity of these strategies to explore different types of networks. We implement the results presented in appendix B to calculate a global time $\tau$ that gives an estimate of the average number of steps needed for a Markovian random walker to reach any destination node for walks with transition probabilities $\pi_{i\to j}$ of Eq. (\ref{wij_g}). The value of $\tau$ is
\begin{equation}\label{TauGeneral}
\tau=\frac{1}{N}\sum_{i=1}^N \tau_i
\end{equation}
where
\begin{equation}
\label{TauiSpect_General}
\tau_i=\sum_{l=2}^N\frac{1}{1-\lambda_l}\frac{\left\langle i|\phi_l\right\rangle \left\langle\bar{\phi}_l|i\right\rangle}{\left\langle i|\phi_1\right\rangle \left\langle\bar{\phi}_1|i\right\rangle}\, .
\end{equation}
In Eq. (\ref{TauiSpect_General}), $\left\langle\bar{\phi}_i\right|$ and $ \left|\phi_i\right\rangle$ with $i=1,2,\ldots, N$ denote the sets of left and right eigenvectors of the transition matrix $\pi_{i\to j}$, these eigenvectors have the respective eigenvalue $\lambda_i$ (see appendix B for details). In the particular case of regular networks  with generalized degree $\mathcal{K}_i=g(\mathbf{L}) _{ii}$ constant for all the nodes, the global time $\tau$ is the Kemeny constant of a Markovian processes
\begin{equation}\label{TauKemeny}
\tau=\sum_{l=2}^N \frac{1}{1-\lambda_l}.
\end{equation}
This quantity only depends on the spectra of the transition matrix and is a result valid for networks with regular generalized degree. In the particular case of normal random walks in a complete graph with $N$ nodes, the value of $\tau$ is $\tau_0=(N-1)^2/N$ (see appendix B).
\\[2mm]
Now, by using the relations in Eqs. (\ref{TauGeneral})-(\ref{TauKemeny}), we explore the random walk strategies defined in Eqs. (\ref{FracDegreeGeneral}), (\ref{gLog}) and (\ref{gExp}) for different types of finite networks. We start our  study with analytical results obtained for finite rings and we continue the analysis for large and small-world networks.
\subsection{Kemeny constant for finite rings}
\begin{figure}[t!]
\begin{center}
\includegraphics[width=0.84\textwidth]{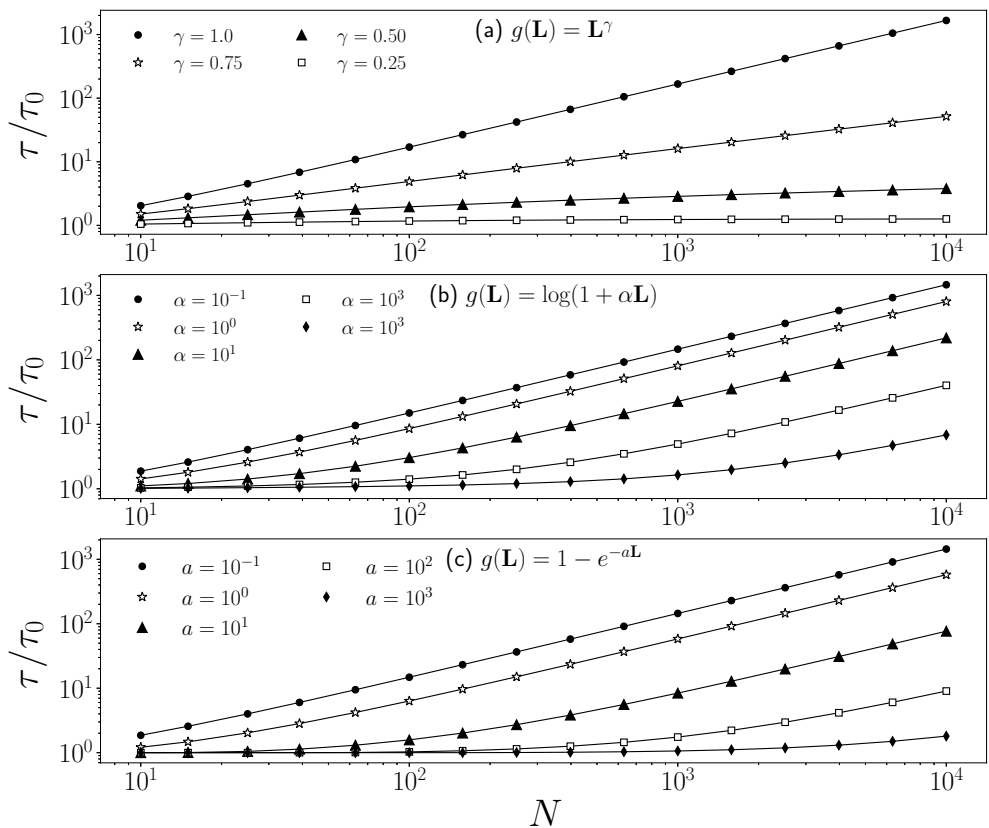}
\end{center}
\vspace{-6mm}
\caption{\label{Fig4} Global time $\tau$ as a function of the number of nodes $N$ for finite rings and different types of random walk strategies defined in terms of function of the Laplacian matrix $\mathbf{L}$. We obtain the results  for the time $\tau$ by direct evaluation of the Eq. (\ref{TauRingEq}) for (a) the fractional Laplacian in Eq. (\ref{Lgamma_mat}), (b) the logarithmic function given by Eq. (\ref{gLog}) and (c) the exponential function in Eq. (\ref{gExp}). We express the values of $\tau$ in relation to the value $\tau_0=(N-1)^2/N$ for different values of the parameter that defines each strategy. Solid lines are used as a guide. }
\end{figure}
Rings are one dimensional lattices with periodic boundary conditions for which the eigenvalues and eigenvectors of the Laplacian matrix are well known (see references \cite{RiascosMateosFD2015,VanMieghem} and appendix A for details). In addition, for this regular structure, the generalized degree is a constant $\mathcal{K}$ given by
\begin{equation}\label{degreeGLring}
\mathcal{K}=g_{ii}(\mathbf{L})=\frac{1}{N}\sum_{l=1}^N g\left(2-2\cos\left[\frac{2\pi}{N}(l-1)\right]\right)
\end{equation}
and the eigenvalues $\{\lambda_i\}_{i=1} ^N$ of the general transition matrix $\mathbf{\Pi}$, with elements  in Eq. (\ref{wij_g}), are given by
\begin{equation}\label{lambdaGLring}
\lambda_i=1-\frac{1}{\mathcal{K}}\, g\left(2-2\cos\left[\frac{2\pi}{N}(i-1)\right]\right).
\end{equation}
As a consequence of the results in Eqs.  (\ref{degreeGLring}) and (\ref{lambdaGLring}), the global time $\tau$ that characterizes the  global performance of the random strategy in Eq. (\ref{wij_g}) to explore a ring is given by the Kemeny constant (see relation in Eq. (\ref{kemenyregular}))
\begin{equation}\label{TauRingEq}
\tau=\frac{1}{N}\sum_{l=1}^N g\left(2-2\cos\phi_l \right)\sum_{m=2}^N\left\{g\left(2-2\cos\phi_m\right)\right\}^{-1},
\end{equation}
where $\phi_i\equiv\frac{2\pi}{N}(i-1)$.
\\[2mm]
In Figure \ref{Fig4} we represent the values of the global time $\tau$ obtained for the fractional, logarithmic and exponential strategies on rings. The results are obtained  by direct evaluation of the result in Eq. (\ref{TauRingEq}).  We explore the effect of the parameters that define each strategy for different values of the size of the ring $N$. 
\\[2mm]
In the case of the fractional random walk on a finite ring, in Figure \ref{Fig4}(a) we observe that the dynamics with $0<\gamma<1$ always improves the capacity to explore the ring in comparison with a normal random walk recovered in the case $\gamma=1$. This effect is observed in the reduction of the time $\tau$ for $\gamma=0.25,\, 0.5$ and $\gamma=0.75$. On the other hand, in the limit $\gamma\to 0$ the dynamics is equivalent to a normal random walker on a fully connected network  allowing, with the same probability, transitions from one node to any site of the ring \cite{RiascosMateosFD2015}, a similar result to this dynamics is also observed for the case $\gamma=0.25$ for all the values of $N$ analyzed. In relation to the strategies defined in terms of logarithms and exponentials, the behavior observed for the normal random walk is recovered for $\alpha\ll 1$ in Figure \ref{Fig4}(b) and in the limit $a\to 0$ in Figure \ref{Fig4}(c). The local dynamics in these limits is a consequence of the results $\log(
x)\approx x$ and $1-e^{-x}\approx x$ valid for small values of $x$ that allow to recover the normal random walk strategy with transition probabilities $\pi^{\mathrm{NRW}}_{i\to j}=\frac{A_{ij}}{k_i}$. In addition, for $\alpha$ and $a$ large we observe that the emergence of long-range displacements reduces the time $\tau$ but only in the limits $\alpha\to \infty$ and $a\to \infty$ the time $\tau=\tau_0$ is obtained for all the values of $N$.
\\[2mm]
Finally, it is important to mention that the effects of the non-local dynamics in all the cases depicted in Figure \ref{Fig4} change the value of $\tau/\tau_0$ in several orders of magnitude with respect to the local case. This effect is significantly marked in large size rings, as it is observed for $N\geq 1000$.
\subsection{Global time $\tau$ for irregular networks}
\begin{figure}[t!]
\begin{center}
\includegraphics[width=0.84\textwidth]{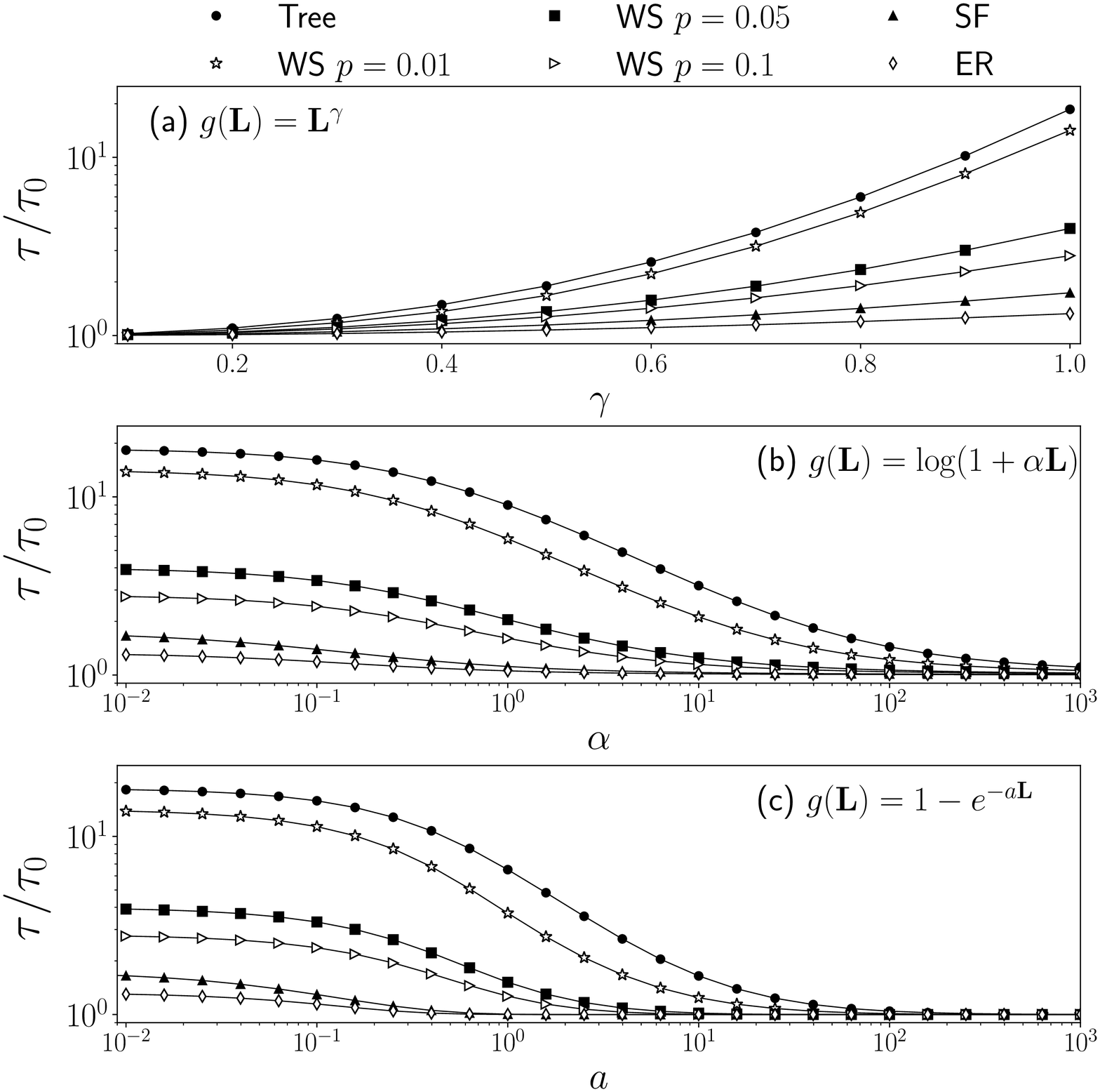}
\end{center}
\vspace{-6mm}
\caption{\label{Fig5} Global time $\tau$ for random walk strategies defined in terms of functions of the Laplacian matrix $\mathbf{L}$ for connected networks with $N=5000$ nodes: a tree, random networks generated from the Watts-Strogatz (WS) model with rewiring probabilities $p=0.01$,  $p=0.05$, $p=0.1$, a scale-free (SF) network of the Barab\'asi-Albert type and a random network of the Erd\H{o}s–R\'enyi (ER) type at the percolation limit $p=\log{N}/N$.  We obtain the results  for the time $\tau$ by numerical evaluation of the Eqs. (\ref{TauGeneral})-(\ref{TauiSpect_General}) for (a) the fractional Laplacian in Eq. (\ref{Lgamma_mat}), (b) the logarithmic function given by Eq. (\ref{gLog}) and (c) the exponential function in Eq. (\ref{gExp}). We express the values of $\tau$ in relation to the value $\tau_0=(N-1)^2/N$ for a fully connected graph. Solid lines are used as a guide. }
\end{figure} 
In this part we analyze the global dynamics of random walkers in different types of large-world and small-world networks. Unlike the previous cases explored for rings, other types of networks have not the same generalized degree $\mathcal{K}_i$ for all the nodes $i=1,2,\ldots,N$. In this way the efficiency or global performance of the random walker is quantified by the time $\tau$ given by the Eq. (\ref{TauGeneral}) that depends of the eigenvectors and eigenvalues of the transition matrix $\mathbf{\Pi}$ with elements given by Eq. (\ref{wij_g}). 
\\[2mm]
In Figure \ref{Fig5} we show the global time $\tau$ for networks with $N=5000$ nodes. We analyze a deterministic tree created by an iterative method for which an initial node ramifies with two leaves that also repeat this process until the size $N$, the final structure is a large-world network with average distances $d$ between nodes that scale as the size of the network. On the other hand, we analyze random networks generated with the Watts-Strogatz model for which an initially regular network is generated and then rewired uniformly randomly with probability $p$; for values of $p\to 0$ this random network has the large world property of the original lattice, however the rewiring introduces shortcuts  that reduce the average path lengths with the increasing of $p$ \cite{WattsStrogatz}. In addition to these, small-world networks generated with the Erd\H{o}s–R\'enyi model and  scale-free (SF) networks of the Barab\'asi-Albert type are explored \cite{ErdosRenyi,BarabasiAlbert}.
\\[2mm]
Once we have the adjacency matrix of each structure, we calculate the respective Laplacian matrix and by using the Eq. (\ref{wij_g}) we calculate the elements of the transition matrix for different functions $g(x)$. Then, through numerical results for the eigenvectors and eigenvalues of the transition matrix we characterize the capacity of each random walk strategy to explore the network by using the time $\tau$ expressed in relation (\ref{TauiSpect_General}) with the value $\tau_0$ for a fully connected graph. Our findings in Figure \ref{Fig5} have similar characteristics to the observed for the case of rings but now, for different structures with the small and large work property. In all the cases studied, we can see that the introduction of long-range displacements always improve the capacity to explore the network in comparison to the observed for the local dynamics that is recovered in the limit cases $\gamma \to 1$ for the fractional dynamics in Figure \ref{Fig5}, $\alpha\to 0$ for the strategy with 
the logarithmic function in Figure 
\ref{Fig5}(b) and  $a\to 0$ for the case defined in terms of exponentials in Figure \ref{Fig5}(c). In the other extreme of a totally non-local dynamics with  $\gamma\to 0$,  $\alpha\gg 1$ and $a\gg 1$, the values of $\tau\to \tau_0$ in agreement with the results observed for the fully connected limit for rings in Figure \ref{Fig4}. 
\\[2mm]
In general we observe that the generalized dynamics defined in terms of the functions $g(\mathbf{L})$ improves the efficiency to explore the networks, the effects are marked in large-world networks with a significant change in the time $\tau$, but the dynamics also improves the results for small-world networks.
\section{Conclusions}
We have deduced conditions that a function of the Laplacian matrix $g(\mathbf{L})$ must satisfy to define a general class of non-local random walks on networks. Examples of these functions are the fractional Laplacian of a graph $\mathbf{L}^{\gamma}$ with $0<\gamma<1$, the logarithmic function  $\log\left(\mathbb{I}+\alpha \mathbf{L}\right)$ for $\alpha>0$ and the function $\mathbb{I}-e^{-a\mathbf{L}}$ with $a>0$. We study the characteristics of the resulting random walks showing that the dynamics is non-local allowing long-range displacements on the network and we obtain analytical results for the transition probabilities for finite and infinite rings. The general formalism introduced contains the fractional random walks as special cases associated to the fractional Laplacian of a graph and L\'evy flights on networks. 
\\[2mm]
We identified two essential types of admissible Laplacian functions  where these two types of good Laplacian functions 
both constitute a certain class of Bernstein functions. Type (i) functions (`Brownian functions') correspond to random walks
with emerging Brownian motion on large networks. Brownian functions have expansions that contain the Laplacian $\mathbf{L}$ as lowest order.
In contrast, there are type (ii) Laplacian functions (`L\'evy functions') with expansions starting with a fractional order $\mathbf{L}^{\gamma}$ ($0<\gamma<1$). Random walks generated by type (ii) functions have on large networks L\'evy flight assymptotics for long-range steps. In both cases {\it the lowest orders of $g(\mathbf{L})$ are the relevant parts} that generate 
the statistics of steps emerging in the limit of large networks $N\rightarrow\infty$: The steps are drawn from Gaussian distributions for walks generated by type (i) functions. In contrast, self-similarly (heavy tailed) distributed long-range steps emerge that are drawn from L\'evy-stable distributions for walks generated by type (ii) functions. For a brief demonstration, we again refer to appendix C.
\\[2mm]
Finally, we evaluate the global capacity of the different random walk strategies defined through the formalism introduced. For the cases explored we identify limits for which the dynamics is reduced to a normal random walk and cases where the jumps between nodes are equivalent to a normal random walk on a fully connected graph. For the cases explored, we conclude that the non-local dynamics generally improves the capacity to visit nodes on the network, a result that is marked in the case of networks with large average distances between nodes like lattices and trees but that also is evident in small-world networks. 
\\[2mm]
We briefly mention a further observation. In the present paper we explore classes of functions which map Laplacians $\mathbf{L}$ on good Laplacians $g(\mathbf{L})$. We can hence define sucessive sequences of good Laplacian functions by the recursion $g^{(n+1)}(\mathbf{L})=g(g^{n}(\mathbf{L}))$ where $g^{(0)}(\mathbf{L})=\mathbf{L}$ and $g^{(n\rightarrow\infty)}(\mathbf{L})$ also constitutes a good Laplacian function which exists only if $g^{(n\rightarrow\infty)}(x) < \infty $ on the spectral interval of $\mathbf{L}$ remains finite. As $g(\mathbf{L})$ in general are non-linear functions of $\mathbf{L}$, there may exist interesting links to fractal maps.
We see that in such an iterative process type (i) functions remain type (i) functions (as their first order term $x$ remains stable) whereas type (ii) functions after $n$ iterations start with lowest order $\mathbf{L}^{\gamma^n}$ (where $0<\gamma^ n < 1$) where $\gamma^n \rightarrow 0+$ for $n\rightarrow\infty$ taking for $n\rightarrow \infty$ an infinitesimally positive exponent approaching zero from the right hand side. This limit of vanishing exponent corresponds to a complete graph (fully connected network) and was analyzed recently \cite{Michelitsch2017PhysARecurrence}.
It seems that in this recursive way we may define random walk strategies which call for further analysis.
\section{Appendix A. Function $g(\mathbf{L})$ for infinite one-dimensional lattices}
In this appendix we explore the form of the function $g(\mathbf{L})$ for rings with $N$ nodes and the limit  $N\to \infty$. In the particular case of rings, the periodicity of the system allows to obtain some useful results in terms of sums and integrals. We follow a similar approach as introduced in references \cite{RiascosMateosFD2014,RiascosMateosFD2015}. 
\\[2mm]
A ring is a one-dimensional lattice with periodic boundary conditions, each node has degree $k=2$.
In this case, the Laplacian is a circulant matrix for which its eigenvectors and eigenvalues can be obtained analytically \cite{VanMieghem}. In particular, the eigenvectors $\{| \Psi_l \rangle\}_{l=1}^N$ of the Laplacian matrix of a circulant network are given by  $\langle m | \Psi_l \rangle=\xi^{(l-1)(m-1)}/\sqrt{N}$  with $\xi\equiv\exp[-\textrm{i}2\pi/N]$, where we denote $\textrm{i}=\sqrt{-1}$ \cite{VanMieghem}. On the other hand, the unsorted eigenvalues of the Laplacian matrix for a ring with $N$ nodes are given by \cite{VanMieghem}
\begin{equation}
\mu_m=2-2\cos\left[\frac{2\pi(m-1)}{N}\right] \qquad \textrm{for} \qquad m=1,\ldots,N.
\end{equation}
By using these results and Eq. (\ref{FunGLap}) we obtain the elements of the function $g(\mathbf{L})$ for a ring with $N$ nodes
\begin{equation}
g_{ij}(\mathbf{L})=\frac{1}{N}\sum_{l=1}^N\, g\left(2-2\cos\left[\frac{2\pi}{N}(l-1)\right]\right) e^{\textrm{i}\frac{2\pi}{N}(l-1)(i-j)}.
\end{equation}
In this relation $e^{\textrm{i}\frac{2\pi}{N}(l-1)(i-j)}=e^{\textrm{i}\frac{2\pi}{N}(l-1)d_{ij}}$, where $d_{ij}$ is the distance between nodes $i$ and $j$ in the ring. This relation reveals directly how the resulting random walk strategy in Eq. (\ref{wij_g}) allows transitions not only to first nearest-neighbors but displacements at any distance. In order to clarify this result we take the limit $N\to \infty$ for which the sum can be approximated by an integral.
\\[2mm]
In the limit $N\to\infty$, the introduction of the variable $\theta=\frac{2\pi}{N}(l-1)$ and the respective differential  $d\theta=\frac{2\pi}{N}$, allows to obtain for an infinite ring
\begin{equation}\label{Lijintegral}
g_{ij}(\mathbf{L})=\frac{1}{2\pi}\int_{0}^{2\pi} g\left(2-2\cos\theta\right) e^{\textrm{i}d_{ij}\theta}d\theta.
\end{equation}
In this way, the transition probability $ \pi_{i \to j}$ defined in (\ref{wij_g}) takes the particular form for an infinite ring
\begin{equation}\label{wijintegral}
\pi_{i \to j}=\delta_{ij}-\frac{1}{2\pi \mathcal{K}}\int_{0}^{2\pi} g\left(2-2\cos\theta\right) e^{\textrm{i}d_{ij}\theta}d\theta,
\end{equation}
where $\mathcal{K}$ is the generalized degree given by
\begin{equation}\label{integralKdegree}
\mathcal{K}=\frac{1}{2\pi}\int_{0}^{2\pi} g\left(2-2\cos\theta\right) d\theta.
\end{equation}
The results in Eqs. (\ref{wijintegral}) and (\ref{integralKdegree}) allow to explore the transition probabilities for different functions $g(x)$. By using numerical integration we obtain the results in Figures \ref{Fig1}-\ref{Fig3} where the non-local character of the random walk strategy introduced in Eq. (\ref{wij_g}) is revealed. In addition, by using the methods developed in \cite{Michelitsch2017PhysA} for the power function, different asymptotic results can be deduced for $n$-dimensional lattices.
\section{Appendix B. Global characterization of random walk strategies}
In this appendix we explore the global time $\tau$ and the Kemeny constant $\mathcal{K}$  of a random walker. These quantities allow to quantify the performance of the random walk strategies defined in Eq. (\ref{wij_g}) to explore a network. Due to the fact that the matrices $g(\mathbf{L})$ are symmetric, i.e. $g_{ij}(\mathbf{L})=g_{ji}(\mathbf{L})$, different quantities that characterize the random walk such as the mean first passage time (MFPT) can be calculated by using a formalism based on random walks on weighted networks \cite{ZZhangPRE2013}. We follow a similar approach as implemented in the study of L\'evy random walks on networks (see reference \cite{RiascosMateosLF2012} for details). 
\\[2mm]
In the case of the random walker with transition probabilities $\pi_{i\to j}$ in Eq. (\ref{wij_g}), we present explicit relations to calculate different quantities in terms of eigenvectors and eigenvalues of the transition matrix $\mathbf{\Pi}$ with elements  $\pi_{i\to j}$. We start with the matrical form of the master equation
\begin{equation}	
\vec{P}(t)=\vec{P}(0)\mathbf{\Pi}^t  \, ,
\end{equation}
here $\vec{P}(t)$ is the probability vector at time $t$. Using Dirac's notation
\begin{equation}\label{ProbVector}
P_{ij}(t)=\left\langle i\right|\mathbf{\Pi}^t \left|j\right\rangle,
\end{equation}
where $\{\left|m\right\rangle \}_{m=1}^N$ represents the canonical base of $\mathbb{R}^N$. Due to the symmetry of $g(\mathbf{L})$, the transition matrix $\mathbf{\Pi}$ can be diagonalized and its spectrum has real values \cite{VKampen}. For right eigenvectors of $\mathbf{\Pi}$ we have $\mathbf{\Pi}\left|\phi_i\right\rangle=\lambda_i\left|\phi_i\right\rangle $ for $i=1,..,N$, where the set of eigenvalues is ordered in the form $\lambda_1=1$ and $1>\lambda_2\geq..\geq\lambda_N\geq -1 $. On the other hand, from right eigenvectors we define a matrix $\mathbf{Z}$ with elements $Z_{ij}=\left\langle i|\phi_j\right\rangle$. The matrix $\mathbf{Z}$ is invertible and, a new set of vectors $\left\langle \bar{\phi}_i\right|$ is obtained by means of $(\mathbf{Z}^{-1})_{ij}=\left\langle \bar{\phi}_i |j\right\rangle $, as consequence
\begin{equation}\label{cond1}
\delta_{ij}=(\mathbf{Z}^{-1}\mathbf{Z})_{ij}=\sum_{l=1}^N \left\langle\bar{\phi}_i|l\right\rangle \left\langle l|\phi_j\right\rangle=\langle\bar{\phi}_i|\phi_j\rangle\, , 
\end{equation}
\begin{equation}\label{cond2}
\mathbb{I}=\mathbf{Z}\mathbf{Z}^{-1}=\sum_{l=1}^N \left|\phi_l\right\rangle \left\langle \bar{\phi}_l \right| \, ,
\end{equation}
where $\mathbb{I}$ is the $N\times N$ identity matrix. Now, by using the diagonal matrix $\mathbf{\Delta} \equiv \textrm{diag}(\lambda_1,\ldots,\lambda_N)$ is obtained $\mathbf{\Pi}=\mathbf{Z}\mathbf{\Delta}\mathbf{Z}^{-1}$, therefore (\ref{ProbVector}) takes the form
\begin{equation}\label{PtSpect}
	P_{ij}(t)=\left\langle i\right|\mathbf{Z}\mathbf{\Delta}^t\mathbf{Z}^{-1}\left|j\right\rangle
	= \sum_{l=1}^N\lambda_{l}^t\left\langle i|\phi_l\right\rangle \left\langle \bar{\phi}_l|j\right\rangle  \, .
\end{equation}
From (\ref{PtSpect}), the stationary distribution $P_i^\infty$ (probability to find the random walker in the  node $i$ in the limit $t\to\infty$) is  $P_j^{\infty}=\left\langle i|\phi_1\right\rangle \left\langle \bar{\phi}_1|j\right\rangle$, where the result $\left\langle i|\phi_1\right\rangle=\mathrm{constant}$ makes $P_j^{\infty}$ independent of the initial condition. On the other hand, the time $\tau_i$ that quantifies the average number of steps needed for the random walker to reach the node $i$ is given by \cite{RiascosMateosLF2012}
\begin{equation}
\label{TauiSpect}
	\tau_i=\sum_{l=2}^N\frac{1}{1-\lambda_l}\frac{\left\langle i|\phi_l\right\rangle \left\langle\bar{\phi}_l|i\right\rangle}{\left\langle i|\phi_1\right\rangle \left\langle\bar{\phi}_1|i\right\rangle}\, ,
\end{equation}
Additionally to this time, for $i \neq j$, we have the MFPT $\left\langle T_{ij}\right\rangle$
\begin{equation}\label{TijSpect}
\left\langle T_{ij}\right\rangle
=\sum_{l=2}^N\frac{1}{1-\lambda_l}\frac{\left\langle j|\phi_l\right\rangle \left\langle\bar{\phi}_l|j\right\rangle-\left\langle i|\phi_l\right\rangle \left\langle\bar{\phi}_l|j\right\rangle}{\left\langle j|\phi_1\right\rangle \left\langle\bar{\phi}_1|j\right\rangle}\, ,
\end{equation}
whereas for $i=j$ $\left\langle T_{ii}\right\rangle=(\left\langle i|\phi_1\right\rangle \left\langle\bar{\phi}_1|i\right\rangle)^{-1}$. Finally, we have the Kemeny's constant
\begin{equation}\label{KconstSpect}
\sum_{m=1}^N\sum_{l=2}^N \frac{1}{1-\lambda_l} \left\langle\bar{\phi}_l|m\right\rangle \left\langle m|\phi_l\right\rangle =\sum_{l=2}^N \frac{1}{1-\lambda_l}
\end{equation}
result that only depends on the spectrum of $\mathbf{\Pi}$. 
Now, we are interested in a global time to describe the global capacity of the random walker to explore a network. We use the global quantity \cite{RiascosMateosLF2012}
\begin{equation}\label{tauglobal}
\tau\equiv\frac{1}{N}\sum_{i=1}^{N}\tau_i \,  ,
\end{equation}
that gives a value associated with the mean time to reach any site of the network. In the particular case of random walks on a special type of regular networks for which the value of the generalized degree $\mathcal{K}_i=-\sum_{l\neq i}g_{il}(\mathbf{L})$ is a constant, the stationary distribution is $P_{i}^{\infty}=1/N$. We denominated these structures {\it generalized regular networks} due to the fact that  according to the result in Eq. (\ref{SumApowers}) is required that the number of nodes at distances $1,2,3\ldots$  is the same for all the nodes in the network, fully connected graphs, rings and simple cubic lattices with periodic boundary conditions are structures with this type of regularity. As a consequence, for generalized regular networks, using Eqs. (\ref{TauiSpect}) and (\ref{tauglobal}) we have for the global time $\tau$
\begin{equation}\label{EffectRegular}
\tau_{\mathrm{reg}}=\sum_{l=2}^N \frac{1}{1-\lambda_l} \, ,
\end{equation}
therefore $\tau_{\mathrm{reg}}$ is equal to the Kemeny's constant where the summation is performed over all $\lambda_l \neq 1$. In view of relation (\ref{wij_g}) there is in generalized regular networks a simple relation between the eigenvalues of the transition matrix $\lambda_m$ and those of $g(\mathbf{L})$, namely
\begin{equation}
\label{eigvalrelation}
\lambda_m = 1-\frac{g(\mu_m)}{{\mathcal K}} , 
\end{equation}
where, as a consequence of Eq. (\ref{AverageK}), the generalized degree is determined by ${\mathcal K} =\frac{1}{N}\sum_{l=2}^Ng(\mu_l)$. For generalized regular networks we thus obtain for the Kemeny constant (\ref{EffectRegular}) the simple expression
\begin{equation}
\label{kemenyregular}
\tau_{\mathrm{reg}}= {\mathcal K} \sum_{l=2}^N\frac{1}{g(\mu_m)} = \frac{1}{N}\sum_{l=2}^Ng(\mu_l)\sum_{l=2}^N\frac{1}{g(\mu_m)} .
\end{equation}
where as mentioned the generalized degree ${\mathcal K}$ is constant for all nodes.
An example of this simplification is given by the normal random walks on a complete graph. This case illustrates the best scenario for the exploration of a network by means of normal random walks since all the nodes are connected. For a complete graph $A_{ij}=1-\delta_{ij}$ and $\pi_{i\to j}=\frac{1-\delta_{ij}}{N-1}$ \cite{VanMieghem}. The eigenvalues of the matrix $\mathbf{\Pi}$ are $\lambda_1=1$ and $\lambda_2=\ldots=\lambda_N=-(N-1)^{-1}$, then the Kemeny's constant (\ref{KconstSpect}) for unbiased random walks on a complete network is
\begin{equation}
\tau_{0}=\frac{(N-1)^2}{N} \, ,
\end{equation}
this is the lowest value that $\tau$ can take.
\section{Appendix C. Asymptotic properties of Laplacian functions}
The aim of this paragraph is to demonstrate the asymptotic relation in Eq. (\ref{case1}), the lowest orders of (\ref{case1}) and their effect on the asymptotic behavior of the transition matrix for long-range steps on sufficiently large networks.
\subsection{Type (i) and (ii) Laplacian functions}
From the results established in section \ref{FunctionSection}, we have seen that Laplacian functions $g(\mathbf{L})$ have to fulfill conditions I, II, III. Let us now analyze 
the behavior of $g(x)$ for $x\rightarrow 0+$ in Eq. (\ref{case1}). Considering  non-negative matrix functions of the Laplacian $\mathbf{L}$ generated by monotonous functions in terms of a series as in Eqs. (\ref{hmatrixBseries})-(\ref{hij_nonnegative}), we have
\begin{eqnarray}\nonumber
f(\mathbf{L}t)&= h(\kappa \mathbb{I} -t\mathbf{L})= \sum_{n=0}^{\infty}\frac{h^{(n)}(0)}{n!}(\kappa \mathbb{I} - t\mathbf{L})^n \\
&= \sum_{n=0}^{\infty}\frac{h^{(n)}(\kappa)}{n!}(-t)^n\mathbf{L}^n , \qquad 0\leq t \leq 1, \label{monotonous}
\end{eqnarray}
where we denote $h^{(n)}(z)\equiv\frac{d^n}{dx^n}h(x)\big|_{x=z}$. Further we have $f(x)=\frac{d}{dx}g(x) >0$, $(-1)^n\frac{d^n}{dx^n}f(x) \geq 0$ ($n=1,2,\ldots$)  and all non-vanishing coefficients $h^{(n)}(0)=(-1)^n\frac{d^n}{dx^n}f(x)|_{x=0}$ are positive. 
\\[2mm]
Generating a monotonously increasing function
$h(\xi) >0 $ and $\xi\geq 0$ with $\kappa > \mu_N$ yields (where $H'(\xi)=\frac{d}{d\xi}H(\xi) = h(\xi) >0 $):
\begin{eqnarray}
\nonumber
\frac{d}{dt}g(t\mathbf{L})\Big|_{t=0} &= f(0)\mathbf{L} = \frac{d}{dt}(H(\kappa)\mathbb{I}-H(\kappa \mathbb{I} - t\mathbf{L}))\Big|_{t=0}\\
&=-\frac{d}{dt}H(\kappa \mathbb{I} - t\mathbf{L})\Big|_{t=0}=  h(\kappa)\mathbf{L} ,\qquad h(\kappa)>0. \label{derivative}
\end{eqnarray}
It follows from the monotony $\frac{d}{d\xi}H(\xi)\big|_{\xi=\kappa}= h(\kappa) >0$  in Eq. (\ref{choice1}), that the first order in $\mathbf{L}$ in a good Laplacian function $g(\mathbf{L})$ is {\it non-vanishing and positive}, especially because of $\frac{d}{dx}g(x)\big|_{x=0}=f(0) > 0$ is positive. This is true for
infinitely often everywhere differentiable $C^{\infty}$ functions when a Taylor series of the form (\ref{monotonous}) exists. 
In this way, good Laplacian functions that are infinitely often differentiable everywhere always are of type (i) (Brownian-) functions.
Since rescaled Laplacian functions have the same transition matrix defined in Eq. (\ref{wij_g}), we can renormalize $g(x) \rightarrow \frac{1}{h(\kappa)}g(x)$ to generate an equivalent Laplacian function having the expansion
\begin{equation}
\label{type-i}
g^{(i)}(x) = x+\sum_{n=2}^{\infty}|g_n|(-1)^{n-1} x^n = x+{\tilde g}(x)
\end{equation}
starting with $x$ as lowest non-vanishing order. The expansion in Eq. (\ref{type-i}), which holds only for type (i) functions, is obtained due to the positive non-vanishing derivative $g'(x=0)= f(x=0) >0$ of the Laplacian
function.
\\[2mm]
We now consider the existence of an asymptotic relation in Eq. (\ref{case1}) for type (ii) (L\'evy-) functions: 
The first observation is that $x^{\gamma}$ for $0<\gamma<1$ is not continuously differentiable in $x=0$ where $f(x\rightarrow 0+) =\gamma x^{\gamma-1}\big|_{x\rightarrow 0+} \rightarrow \infty$.
It is hence not possible to generate a type (ii) function
by a Taylor series around $x=0$. We see here that we need to find another way to obtain the lowest non-vanishing order of a type (ii) function.
\\[2mm]
In order to prove Eq. (\ref{case1}) for type (ii) functions let us first consider the Mellin transform
of a good Laplacian type (i) function $g(\mathbf{L})=\mathbb{I}-e^{-\mathbf{L}t}$ with $t>0$
\begin{equation}
\label{matint}
g_{\gamma}(\mathbf{L})= -\frac{1}{\Gamma(-\gamma)}\int_0^{\infty} (\mathbb{I}-e^{-\mathbf{L}t})  t^{-1-\gamma} {\rm d}t ,\qquad 0<\gamma <1.
\end{equation}
We notice that the good properties I, II, III of the Laplacian function $\mathbb{I}-e^{-\mathbf{L}t}$ are maintained by the integral in Eq. (\ref{matint}).  The factor
$-\frac{1}{\Gamma(-\gamma)}=\frac{\gamma}{\Gamma(1-\gamma)} >0$ is a positive normalization constant which is justified below.
Let us now analyze the Mellin transform 
\begin{equation}
\label{matint2gen}
g_{\gamma}^{(ii)}(\mathbf{L})= C_{\gamma} \int_0^{\infty} g^{(i)}(t\mathbf{L}) t^{-1-\gamma} {\rm d}t ,\hspace{0.5cm} 0<\gamma <1,
\end{equation}
where $g^{(i)}(x)$ denotes a good Laplacian type (i) function having an expansion of the general form given by Eq. (\ref{type-i}) and $C_{\gamma} >0$ denotes a positive normalization constant.
In view of Eq. (\ref{type-i}) we see that convergence of Eq. (\ref{matint2gen}) for $t\rightarrow 0$ requires, due to the asymptotic behavior of $g^{(i)}(x) \approx x$ ($x\rightarrow 0$), exponents $\gamma <1$.
On the other hand, the convergence of Eq. (\ref{matint2gen}) is fulfilled if the Mellin transform
\begin{equation}
 \label{mellinexistence}
 M_g^{(i)}(-\gamma) =  \int_0^{\infty} g^{(i)}(t) t^{-1-\gamma} {\rm d}t
\end{equation}
exists for functions $g^{(i)}(x)$ which do not increase more rapidly than $\log(x)$ including type (i) functions that approach a constant value for $x\rightarrow \infty$ such as $1-e^{-x} \rightarrow 1$.
For this category of type (i) functions $\gamma >0$ is required in Eq. (\ref{mellinexistence})
thus $0<\gamma< 1$ guarantees then existence of the Mellin transform (\ref{mellinexistence}). We notice that no other type (i) function is admitted in (\ref{mellinexistence}).
For instance the good Laplacian function (trivial case)
$g^{(i)}(x) =x$ is not admitted and yields a divergence in Eq. (\ref{mellinexistence}) at $t\rightarrow \infty$ for $\gamma < 1$ whereas $\gamma <1$ is required to avoid a divergence at $t\rightarrow 0+$. 
\\[2mm]
Now, by using $\mathbb{I}-e^{-\mathbf{L}t} = \sum_{m=2}^N(1-e^{-\mu_mt})|\Psi_m\rangle\langle\Psi_m|$, partial integration of the Eq. (\ref{matint}) yields
\begin{eqnarray}\nonumber
g_{\gamma}(\mathbf{L})=& -\frac{1}{\Gamma(-\gamma)}\sum_{m=2}^{\infty}|\Psi_m\rangle\langle\Psi_m| (1-e^{-\mu_mt})\frac{t^{-\gamma}}{(-\gamma)}\Bigg|_0^{\infty}\\
&+\frac{1}{\Gamma(1-\gamma)}\sum_{m=2}^{\infty}|\Psi_m\rangle\langle\Psi_m| 
 \mu_m^{\gamma} \int_0^{\infty}e^{-s}s^{-\mu_m}{\rm d}s.  \label{partint}
\end{eqnarray}
Thus we obtain for $0< \gamma<1$
\begin{eqnarray}\nonumber
g_{\gamma}(\mathbf{L}) &= \frac{1}{\Gamma(1-\gamma)} \int_0^{\infty}t^{-\gamma} \frac{d}{dt}(\mathbb{I}-e^{-\mathbf{L}t}){\rm d}t = {\cal D}_0^{\gamma}(\infty)(\mathbb{I}-e^{-\mathbf{L}t}) \\
&= \frac{1}{\Gamma(1-\gamma)} \int_0^{\infty}t^{-\gamma} e^{-\mathbf{L}t}\mathbf{L} {\rm d}t =     \mathbf{L}^{\gamma},  \label{second}
\end{eqnarray}
where in the first line of this relation ${\cal D}_x^{\gamma}(y)$ denotes the Caputo fractional derivative operator \cite{caputo}.
The result indeed is the fractional power of the Laplacian matrix $\mathbf{L}$ with the restriction $0<\gamma<1$. In this way, we have the representation
\begin{equation}
\label{matintb}
\mathbf{L}^{\gamma} = -\frac{1}{\Gamma(-\gamma)}\int_0^{\infty} (\mathbb{I}-e^{-\mathbf{L}t})  t^{-1-\gamma} {\rm d}t 
\end{equation}
with $0<\gamma <1$ maintaining the good properties I, II, III. 
Now let us return to the general case presented in the integral in Eq. (\ref{matint2gen}). By using the Mellin transform $ M_g^{(i)}(-\gamma)$ of a type (I) function $g^{(i)}(x)$ defined in Eq. (\ref{mellinexistence}) that exists for $0<\gamma<1$ and, by choosing the normalization constant $C_{\gamma}= \frac{1}{M_g^{(i)}(-\gamma)}$ we can write Eq. (\ref{matint2gen}) in the following form
\begin{eqnarray}
\nonumber
g_{\gamma}^{(ii)}(\mathbf{L}) &= \frac{1}{M_g^{(i)}(-\gamma)}\int_0^{\infty} g^{(i)}(t\mathbf{L}) t^{-1-\gamma} {\rm d}t \\ 
&= \frac{1}{M_g^{(i)}(-\gamma)}
\sum_{m=2}^{\infty}|\Psi_m\rangle\langle\Psi_m| \int_0^{\infty}  g^{(i)}(\mu_mt) t^{-1-\gamma} {\rm d}t. \label{matint2gen3}
\end{eqnarray}
We see that the integral in the spectral sum by putting $s=\mu_mt$ takes the form \newline $ \int_0^{\infty}  g^{(i)}(\mu_mt) t^{-1-\gamma} {\rm d}t= \mu_m^{\gamma}  
M_g^{(i)}(-\gamma)$ thus Eq. (\ref{matint2gen3}) becomes
\begin{equation}
\label{matint2gen4}
g_{\gamma}^{(ii)}(\mathbf{L}) = \sum_{m=2}^{\infty}|\Psi_m\rangle\langle\Psi_m| \mu_m^{\gamma}  =\mathbf{L}^{\gamma}.
\end{equation}
Therefore, we have the general result
\begin{equation}
\label{matint2gengen}
g_{\gamma}^{(ii)}(\mathbf{L}) = \frac{1}{M_g^{(i)}(-\gamma)}\int_0^{\infty} g^{(i)}(t\mathbf{L}) t^{-1-\gamma} {\rm d}t =\mathbf{L}^{\gamma} , \qquad  0 < \gamma < 1
\end{equation}
that allows to generate the fractional Laplacian (which is a type (ii) function) in terms of type (i) Laplacian functions
$g^{(i)}(x)$ for which a Mellin transform $M_g^{(i)}(-\gamma) = \int_0^{\infty} g^{(i)}(s) s^{-1-\gamma}{\rm d}s$ exists. The Mellin transformation in Eq. (\ref{matint2gengen})
conserves the good properties I, II, III of the type (i) Laplacian function $g^{(i)}(t\mathbf{L})$ used in Eq. (\ref{matint2gengen}).
\\[2mm]
From these results and Eqs. (\ref{choice1}), (\ref{choice2}) we observe that type (ii) (L\'evy) functions 
$g^{(ii)}(\mathbf{L})=\mathbf{L}^{\gamma} +{\tilde g}(\mathbf{L})$ have always as lowest non-vanishing orders $\mathbf{L}^{\gamma}$ with $0<\gamma<1$. We notice that functions $x^{\beta}+ {\tilde g}(x)$ with $\beta >1$ (and where
${\tilde g}$ contains only orders greater than $\beta$)
do not fulfill Eq. (\ref{choice2}) and thus are not admissible Laplacian functions. It follows that the only admissible type (ii) functions conserving conditions I,II,III
are functions of the form
\begin{equation}
\label{type-ii}
g^{(ii)}(\mathbf{L}) =\mathbf{L}^{\gamma} + {\tilde g}(\mathbf{L}).
\end{equation}
Functions with $\gamma >1$ cannot be generated by Eq. (\ref{matint2gengen}) since such functions produce a divergent integral (\ref{matint2gengen}) at $t=0$.
In this way we have demonstrated the asymptotic relation in Eq. (\ref{case1}) with lowest non-vanishing orders uniquely either $\mathbf{L}$ for type (i) functions, and $\mathbf{L}^{\gamma}$ ($0<\gamma<1$) 
for type (ii) Laplacian functions where only these lowest orders are relevant for the statistics of long-range steps.
\\[2mm]
Also type (ii) functions including linear combinations of powers of the Laplacian are admissible, in this case  $g^{(ii)}(\mathbf{L})  = \sum_{m=1}^S \mathcal{A}_{\gamma_m}\mathbf{L}^{\gamma_m}$ where all exponents satisfy $0<\gamma_m<1$ for $m=1,2,\ldots,S$. In this particular class of functions, the smallest exponent $\gamma_m$ determines the asymptotic behavior for long-range steps. In the following part, we will consider the asymptotics of the transition matrix for these two categories of type (i) and (ii) emerging on large networks.
\subsection{Asymptotic behavior of the transition matrix for type (i) and (ii) Laplacian functions}
In order to discuss the asymptotic behavior of type (i) and type (ii) Laplacian functions we consider the time-discrete random walks defined by the transition probability matrix as in Eq. (\ref{wij_g})
\begin{equation}
 \label{transition}
\pi_{i\rightarrow j} = \delta_{ij} -\frac{1}{\mathcal{K}_i}g_{ij}(\mathbf{L})
\end{equation}
with the generalized degree $\mathcal{K}_i=  g_{ii}{(\bf L})=-\sum_{j\neq i} g_{ij}({\bf L})  >0 $.  Now, for the sake of simplicity of our demonstration we consider here regular undirected networks with constant generalized degree ${\mathcal{K}}_i={\mathcal{K}}$ for all nodes $i=1,\ldots,N$. Such regular networks include for instance
simple cubic $d$-lattices. In this case the transition probability matrix is symmetric $\pi_{i\rightarrow j} =\pi_{ij}=\pi_{ji}$ having the representation
\begin{equation}
\label{theform}
{\bf \Pi} = \mathbb{I}-\frac{1}{\mathcal{K}}g({\bf L}) = \sum_{m=1}^N \lambda_m |\Psi_m\rangle\langle\Psi_m|
\end{equation}
with the same eigenvectors as the Laplacian matrix and Laplacian matrix function with eigenvalues
\begin{equation}
 \label{pims}
 \lambda_m=1-\frac{g(\mu_m)}{\mathcal{K}}
\end{equation}
where $|1-\frac{g(\mu_m)}{\mathcal{K}}| < 1$ for $m=2,\ldots,N$ and the eigenvalue $\lambda_1=1$ is always conserved as a consequence of the property $g(\mu_1=0)=0$.
Further we notice that in such regular networks the constant degree ${\mathcal{K}}$ is completely determined by the trace of the Laplacian matrix function $g(\mathbf{L})$, thus
\begin{equation}
\label{constantdegree}
{\mathcal{K}} = \frac{1}{N}\sum_{m=2}^Ng(\mu_m).
\end{equation}
The regular networks under consideration here topologically correspond, as mentioned above, 
to a $d$-dimensional simple cubic lattice ($d=1,2,3,\ldots$) where we assume infinite boundary conditions in any dimension $j=1,\ldots ,d$ where the number of nodes is infinity ($N\rightarrow\infty$)
to capture emerging asymptotic behavior for probability of long-range steps.
The nodes of this network are represented by the lattice vectors $\vec{p}=(p_1,\ldots ,p_d)$ where the components 
$p_j=0, \pm 1,\pm 2,\ldots  \pm \infty \in \mathbb{Z}$ may take any integer value. The Laplacian matrix ${\cal L}(\vec{p}-\vec{q})$ in this lattice with only next neighbor connections have the matrix elements ${\cal L}_{p_1,\ldots p_n|q_1,\ldots ,q_n}$ given by \cite{Michelitsch2017PhysARecurrence}
\begin{eqnarray} 
{\cal L}_{p_1,\ldots p_n|q_1,\ldots ,q_n} &= 2d \prod_{j=1}^d\delta_{p_jq_j} - \sum_{j=1}^d\left(\delta_{p_{j+1}q_j}+ \delta_{p_{j-1}q_j} \right)\prod_{s\neq j}^n \delta_{p_sq_s} \\ \label{explicit}
&= \frac{1}{(2\pi)^d} \int_{\vec{\kappa}} e^{\textrm{i}\vec{\kappa}\cdot(\vec{p}-\vec{q})}\mu(\vec{\kappa}){\rm d}^q{\vec{\kappa}}, 
\end{eqnarray}
where $\mu(\vec{\kappa}) = 2d-2\sum_{j=1}^d\cos{(\kappa_j)}$ with $-\pi \leq \kappa_j\leq \pi$  for $j=1,\ldots ,d$ and the constant degree $K=2d$ indicates the number of adjacent nodes. The relation in Eq. (\ref{explicit}) indicates the spectral representation in terms of $2\pi$-periodic Bloch eigenfunctions 
$\frac{1}{\sqrt{2\pi}}e^{\textrm{i}\kappa_jp_j}$ 
where
$\vec{\kappa}=(\kappa_1,\ldots ,\kappa_j,\ldots ,\kappa_d)$. In addition, in Eq. (\ref{explicit}) we have introduced the abbreviation
\begin{equation*}
\int_{\vec{\kappa}}h(\vec{\kappa}\cdot(\vec{p}-\vec{q}) ) {\rm d}\vec{\kappa} = \int_{-\pi}^{\pi}{\rm d}\kappa_1 \ldots  \int_{-\pi}^{\pi}{\rm d}\kappa_d h(\kappa_1(p_1-q_1)+\ldots +\kappa_d(p_d-q_d))
\end{equation*}
which indicates integration over the $d$-dimensional first Brillouin zone $(2\pi)^d$.
Further we introduced the unity matrix
$\delta_{pq}\rightarrow \delta_{\vec{p}-\vec{q}} = \prod_{j=1}^d\delta_{p_jq_j}$ of the $d$-dimensional lattice. 
\\[2mm]
Now let us consider the asymptotic power law behavior of the eigenvalues of the Laplacian function for $|{\vec{\kappa}}|\rightarrow 0$. In this limit, the eigenvalues of the Laplacian ${\cal L}$ in Eq. (\ref{explicit}) take the form $\mu(\vec{\kappa}) \approx |\vec{\kappa}|^2$ for the lowest order, thus the eigenvalues of the transition probability matrix in Eq. (\ref{pims}) take
for $|\vec{\kappa}|\rightarrow 0$ the representation
\begin{equation}
 \label{pimsasymptotics}
 \lambda(\vec{\kappa}) \approx 1-\frac{g(|\vec{\kappa}|^2)}{\mathcal{K}}
\end{equation}
and because of asymptotic relation (\ref{case1}) given by $g(x)=x^{\gamma}+{\tilde g}(x)$ with $\gamma=1$ for type (i) and $0<\gamma<1$ for type (ii) functions, the eigenvalues  $\lambda(\vec{\kappa})$ determined by Eq.  (\ref{pimsasymptotics}) write in the lowest orders in $\vec{\kappa}$ as
\begin{equation}
 \lambda(\vec{\kappa}) \approx 1-\frac{|\vec{\kappa}|^{2\gamma}}{\mathcal{K}} 
\end{equation}
thus $\lambda^{(i)}(\vec{\kappa}) = 1-\frac{|\vec{\kappa}|^{2}}{\mathcal{K}}$ for type (i), and $\lambda^{(ii)}(\vec{\kappa})=1-\frac{|\vec{\kappa}|^{2\gamma}}{\mathcal{K}}$  ($0<\gamma<1$) for type (ii) functions. Let us now consider $t>>1$ time steps
and account for $1-\frac{|\vec{\kappa}|^{2\gamma}}{\mathcal{K}} \approx \exp{(-\frac{|\vec{\kappa}|^{2\gamma}}{\mathcal{K}})}$. Then the time evolution of the transition matrix for
$t >>1$ can be written as
\begin{eqnarray}
 \label{timeevol}
\mathbf{\Pi}_{\vec{p}-\vec{q}} (t)& = \frac{1}{(2\pi)^d} \int_{\vec{\kappa}} e^{\textrm{i}\vec{\kappa}\cdot(\vec{p}-\vec{q})} (\pi(\vec{\kappa}))^t {\rm d}\vec{\kappa} \\
&=\frac{1}{(2\pi)^d} \int_{\vec{\kappa}} e^{\textrm{i}\vec{\kappa}\cdot(\vec{p}-\vec{q})} \exp{\left( \frac{ -t|\vec{\kappa}|^{2\gamma}}{\mathcal{K}}\right)} {\rm d}\vec{\kappa} 
\end{eqnarray}
which is for type (i) functions ($\gamma=1$) a Gaussian, and for type (ii) functions ($0<\gamma<1$) a L\'evy stable  heavy tailed distribution with L\'evy index $\alpha=2\gamma$ leading to a heavy tailed probability $\sim |\vec{p}-\vec{q}|^{-d-2\gamma}$ for long-range steps for random walks generated by type (ii) Laplacian functions. Further explicit evaluations and special cases for simple cubic lattices can be found in reference \cite{Michelitsch2017PhysARecurrence}.

\section{References}
\providecommand{\noopsort}[1]{}\providecommand{\singleletter}[1]{#1}%
\providecommand{\newblock}{}

\end{document}